\documentclass[12pt,preprint]{aastex}

\usepackage{natbib}
\slugcomment{Accepted for publication in Astrophys J.}

\shorttitle{Dynamics and neutrino signal of black hole formation}
\shortauthors{Sumiyoshi et al.}

\begin{document}


\title{Dynamics and neutrino signal of black hole formation \\
in non-rotating failed supernovae. I. EOS dependence}


\author{K. Sumiyoshi}
\affil{Numazu College of Technology, 
       Ooka 3600, Numazu, Shizuoka 410-8501, Japan}
\email{sumi@numazu-ct.ac.jp}

\author{S. Yamada}
\affil{Science and Engineering
       \& 
       Advanced Research Institute for Science and Engineering, \\
       Waseda University, 
       Okubo, 3-4-1, Shinjuku, Tokyo 169-8555, Japan}

\and

\author{H. Suzuki}
\affil{Faculty of Science and Technology, Tokyo University of Science,\\
       Yamazaki 2641, Noda, Chiba 278-8510, Japan}




\begin{abstract}
We study the black hole formation and the neutrino signal 
from the gravitational collapse of a non-rotating massive star of 40 M$_\odot$.  
Adopting two different sets of realistic equation of state (EOS) of dense matter, 
we perform the numerical simulations of general relativistic $\nu$-radiation hydrodynamics 
under the spherical symmetry.  
We make comparisons of the core bounce, the shock propagation, 
the evolution of nascent proto-neutron star and the resulting 
re-collapse to black hole to reveal the influence of EOS.  
We also explore the influence of EOS on the neutrino emission 
during the evolution toward the black hole formation.  
We find that the speed of contraction of the nascent proto-neutron star, 
whose mass increases fast due to the intense accretion, is different 
depending on the EOS and the resulting profiles of density and temperature 
differ significantly.  
The black hole formation occurs at 0.6--1.3 sec after bounce 
when the proto-neutron star exceeds its maximum mass, 
which is crucially determined by the EOS.  
We find that the average energies of neutrinos increase 
after bounce because of rapid temperature increase,
but at different speeds depending on the EOS.  
The duration of neutrino emission up to the black hole formation 
is found different according to the different timing of re-collapse.  
These characteristics of neutrino signatures are distinguishable from those for 
ordinary proto-neutron stars in successful core-collapse supernovae.  
We discuss that a future detection of neutrinos from black-hole-forming 
collapse will contribute to reveal the black hole formation and 
to constrain the EOS at high density and temperature.  
\end{abstract}


\keywords{supernovae: general --- stars: neutron --- black hole physics --- 
neutrinos --- hydrodynamics --- equation of state}

\section{Introduction}
Massive stars having more than $\sim$10 solar masses (M$_\odot$) end 
their lives when the iron core is formed after the stages 
of nuclear burning \citep{bet90}.  
For the massive stars of $\sim$10--20M$_\odot$, the spectacular events 
known as supernova explosions occur due to the gravitational 
collapse of iron core and the launch of shock wave by the 
bounce of core at high density.  
A successful explosion leaves a proto-neutron star 
which emits a bunch of neutrinos during the formation and cools 
down \citep{bur88,suz94}.  
This duration, ($\sim$20 s), of supernova neutrinos is determined 
by the time scale of diffusion of neutrinos in dense matter
inside the proto-neutron star.  
In the case of SN 1987A, the burst of neutrinos was detected 
at the terrestrial detectors \citep{bio87,hir87} 
and has proved the general scenario 
of supernova explosion and the formation of dense compact object \citep{sat87}.  

Stars more massive than $\sim$20M$_\odot$ may have different fates.  
They usually have larger iron cores and 
will be intrinsically too massive to have stellar explosion.  
Then, the outcome will be a formation of black hole 
regardless of detailed scenarios.  
Although the number of such black-hole-forming massive stars is uncertain 
depending on the mass range and the initial mass function, 
it definitely consists of a portion of the stellar mass 
distribution and can be a substantial fraction.  
Therefore, it is important to clarify the general feature of 
black-hole-forming phenomena in a broader context of 
the gravitational collapse of massive stars.
It is especially exciting to predict 
the neutrino signals from this type of events 
as a clear and unique identification of the black hole formation. 
The calculated template of neutrino signal will be decisive 
to identify the black hole formation in the Galaxy, 
if occurs, by the terrestrial neutrino research facilities.  

Astronomically, 
the fate of massive stars beyond the mass limit ($\sim$20M$_\odot$)
for ordinary supernovae is currently attracting interest \citep{heg03}.  
Recent analyses of the observed light curves of supernovae 
with the explosive nucleosynthesis suggest 
that there are novel categories of explosive phenomena with 
kinetic energies and ejected $^{56}$Ni mass being different from 
those for the ordinary supernovae \citep{mae03}.  
In fact, 
faint supernovae and hypernovae might be the 
outcomes of massive stars in the range of $\sim$20--60M$_\odot$.  
They differ from each other probably in the degree of 
rotation and fall-back, and are supposed to form a black hole 
regardless 
and are clearly distinguished from the ordinary supernovae 
forming neutron stars.
Gamma ray bursts are thought to be commonly associated with 
the hypernovae (collapsars) \citep{mac99}.  
The survey of the fate as a function of mass and metallicity is important 
to understand the evolution of galaxies and the Universe.

The scenarios for the formation of stellar mass black hole 
have been argued in several different contexts 
and the associated neutrino bursts have been studied 
in some cases.  
In the case of successful explosions from the massive stars 
of $\sim$10--20M$_\odot$, some of the proto-neutron stars after the birth 
may evolve toward the black hole even after the launch of shock wave. 

In most of supernova explosions, the accretion of material 
ceases within several 100 msec after bounce and a proto-neutron 
star is born with a fixed total baryon mass.  
Since the maximum mass of proto-neutron stars, which are hot 
and lepton-rich, can be larger than the maximum mass of cold 
neutron stars, massive proto-neutron stars in a certain 
mass range can lead to the black hole formation during the cooling 
and deleptonization \citep{gle95}.  
The scenario of metastable proto-neutron stars relies on 
the equation of state of dense matter having the appearance 
of exotica such as 
hyperons, kaon condensates and quarks during the cooling 
through neutrino emissions \citep{kei95,pon99,pon01b,pon01a}.  
The emergence of neutrinos diffusing out from proto-neutron 
stars has been calculated in the flux-limited diffusion 
approximation during the quasi-hydrostatic evolution \citep{pon99}.  
In the case of metastable proto-neutron stars, 
the termination of neutrino burst is predicted to occur 
in 1 sec $-$ 100 sec, which largely depends on the internal 
composition.  
The disappearance occurs during the exponential decrease of 
neutrino luminosities, which is common in the long-term proto-neutron star 
cooling.  
The delayed collapse of a proto-neutron star in this context 
has been studied in the dynamical simulations with general 
relativistic treatment to follow the final moment of black 
hole formation and the associated neutrino signals 
by the leakage or diffusion scheme \citep{bau96b,bau96a}.  

The collapse to the black hole can be triggered 
if there is a significant fall back of material after 
the successful supernova explosion.  
The fall back in supernovae is used as a given ingredient 
in the studies of explosive nucleosynthesis, but has not 
been well established in the context of explosion mechanism.  
The amount and time scale of fall back is uncertain 
especially in the numerical simulations of supernova 
explosion, whose puzzle has not been solved yet.  
In the aim of examining the supernova neutrinos from SN1987A, 
the quasi-static evolutions of proto-neutron star 
leading in some cases to the black hole formation 
have been studied for the assumed accretion rate \citep{bur88}.  
The numerical simulations have been done by solving the 
quasi-static evolution with the flux-limited diffusion 
approximation, therefore, the general relativistic 
instability was assumed to occur when the numerical code can not 
find a static solution.  
The characteristics of neutrino signals in the case of 
black hole formation were pointed out with the uncertainties 
in accretion rate and equation of state.  

The current study focuses on the black hole formation 
without supernova explosion from massive stars of $\gtrsim$20M$_\odot$, 
being different from the delayed collapse of proto-neutron star 
in successful explosions discussed above.
Having large iron cores, the shock wave can not propagate outward 
and the standing accretion shock is settled above the proto-neutron star.  
The material of outer layers falls down continuously from 
the beginning of gravitational collapse.  
The mass of proto-neutron star increases accordingly toward 
the maximum mass and the dynamical collapse to the black 
hole takes place.  
There is no bright optical display associated with the collapse accordingly.

We aim to clarify the sequence of evolution starting 
from the beginning of gravitational collapse of a massive star of 40M$_\odot$ 
to the final moment of black hole formation.  
In order to predict the time profile of neutrino burst and the energy 
spectrum of neutrinos as well as the dynamical evolution, we perform 
elaborate simulations of $\nu$-radiation hydrodynamics in general 
relativity.  
We follow the evolution of proto-neutron star with matter accretion 
for a long time ($\sim$1 s), which is challenging numerically with the exact 
treatment of $\nu$-radiation hydrodynamics, 
to find when and how the black hole is formed.  

The explosion energies and their dependence on progenitor models 
have been studied by two-dimensional simulations 
of hydrodynamics with a simplified neutrino treatment 
for 15, 25 and 40M$_\odot$ stars \citep{fry99}.
Although the border line for black hole formation without explosion 
has been estimated to be roughly 40M$_\odot$, 
the simplified treatment of neutrinos has allowed the author to provide 
only an approximate limit for success of explosion and 
the evolution of proto-neutron star toward black hole has not been 
studied in his simulations.  
A simulation of general relativistic $\nu$-radiation hydrodynamics 
has been first done by \citet{lie04} and 
the numerical result for a 40M$_\odot$ star leading to the black hole formation 
has been reported with the detailed description of numerical code 
and test bed for core-collapse supernovae but for a single EOS.  

Rapid formations of black hole after the birth of proto-neutron star 
can be crucially affected by the equation of state (EOS) of dense matter.  
The maximum mass of evolving proto-neutron stars is determined 
by the stiffness of EOS and the degree of deleptonization.  
In addition, the initial profile of proto-neutron star is determined 
right after the core bounce, which is influenced by the EOS also.  
Recently, different EOS's for supernova simulations have become 
available \citep{lat91,she98a} and the detailed comparisons of 
core collapse from a 15M$_\odot$ star, which is supposed 
to explode forming a neutron star, have been made in the $\nu$-radiation 
hydrodynamics \citep{sum05}.  
It has become apparent that the profiles of proto-neturon star 
long after bounce are significantly different depending on the EOS 
and the resulting spectra of neutrinos are distinct.  
Therefore, it is essential to examine the influence of the EOS 
on the evolution toward the black hole formation and to predict 
the associated neutrino signals.  
We note again that the single EOS \citep{lat91} has been adopted 
in the previous study by \citet{lie04} and the crucial influence 
of EOS has not been explored yet.

In the current study, we make the first comparison of the dynamical 
formations of black hole from a massive star 
with $\nu$-radiation hydrodynamics by adopting two sets of EOS.  
We aim to clarify the difference of dynamical evolutions  
due to the EOS's in order to probe the dense matter.  
In the black-hole-forming collapse, 
the central core experiences higher densities and temperatures 
than in normal supernovae as we will show.  
It is especially interesting to reveal the differences 
in neutrino signals in order to extract the information of EOS 
from terrestrial neutrino detections in future.  
We will argue that the duration of neutrino burst 
from the black-hole-forming collapse 
can put a constraint on the stiffness of EOS 
and the phase transition point to exotic phases \citep{sum06}.  

We arrange this article as follows.  
We begin with the descriptions on the numerical simulations 
in $\S$ \ref{sec:simulations}, including the numerical code, 
the physical inputs, the initial setting and the criterion for 
black hole formation.  
We report in order 
the numerical simulations in $\S$ \ref{sec:results}, 
describing the evolution of core-collapse, proto-neutron star 
evolution and the second collapse to black hole.  
The evolution of neutrino distributions and 
EOS-dependences in neutrino signals are presented 
in $\S$ \ref{sec:nu-distributions} and \ref{sec:neutrino}, respectively.  
Implications of current results for the EOS study are discussed 
in $\S$ \ref{sec:EOS-implications}.
We end up with a summary in $\S$ \ref{sec:summary}.  

\section{Numerical Simulations}\label{sec:simulations}

\subsection{Numerical Code}
We adopt the numerical code 
of general relativistic $\nu$-radiation hydrodynamics 
that solves the Boltzmann equation for neutrinos 
together with lagrangian hydrodynamics 
under the spherical symmetry \citep{yam97,yam99,sum05}.  
The numerical code has been successfully applied 
to study the core collapse 
in the case of a 15M$_\odot$ star as the progenitor \citep{sum05}.  
It is a fully implicit code in order to follow 
the long-term evolution of supernova core for more than 1 s 
after bounce 
and has a rezoning feature to efficiently capture 
the accretion of outer layer onto the compact object.
The same numerical treatment is applied to the case of 
a 40M$_\odot$ star.  
We adopt 255 spatial zones for lagrangian mass coordinate and 
14 energy zones and 6 angle zones for the neutrino 
distribution function.  
The 4 species of neutrinos are treated separately 
as $\nu_e$, $\bar{\nu}_e$, $\nu_{\mu/\tau}$ 
and $\bar{\nu}_{\mu/\tau}$.
Here $\nu_{\mu/\tau}$ denotes the neutrinos of $\mu$- and $\tau$-types 
and $\bar{\nu}_{\mu/\tau}$ denotes 
the anti-neutrinos of $\mu$- and $\tau$-types.  
The other details of methods can be found in \citep{sum05}.  

We remark that the exact treatment of $\nu$-radiation hydrodynamics 
in general relativity is essential to describe the evolution 
including compact objects and to predict the neutrino emission 
in the dynamical situation.  
The general relativity plays a dominant role to trigger 
the collapse of proto-neutron star to black hole.  
The hydrodynamical treatment is necessary to describe the 
accretion of material onto a shrinking central core 
and the neutrinos mainly emitted during this accretion phase 
being different from the quasi-static evolution of 
proto-neutron stars by the Henyey-type treatment.  

The exact treatment of general relativistic $\nu$-radiation hydrodynamics 
is possible only under the spherical symmetry 
at the moment \citep[See, however,][for efforts on multi-dimensional calculations]{bura03,bura06,liv04}.  
We note that the spherically symmetric models considered in this paper 
may correspond to the branch extending from 
the faint supernovae that are suggested to be the outcome of 
the collapse of slowly rotating massive stars \citep{nom05}.  
On the other hand, 
multi-dimensional simulations of $\nu$-radiation hydrodynamics 
are required to reveal the black hole formation in rapidly rotating 
massive stars and its possible consequence as hypernovae and/or 
gamma ray bursts, 
which are, however, beyond the scope of the current study.
%

\subsection{Input Physics}
\subsubsection{Equation of state}\label{sec:EOS}
We use the two sets of supernova EOS which are the standard 
in recent supernova simulations.  
The EOS by Lattimer and Swesty (LS-EOS) \citep{lat91},
which has been a conventional choice, 
is based on the compressible liquid drop model for 
nuclei together with dripped nucleons.  
The density dependence of EOS for infinite matter 
is assumed to take a form for the Skyrme-type effective interactions, 
which are often used in non-relativistic nuclear many body calculations
of nuclear structures.  
The parameters are determined by the bulk properties at the saturation.
The EOS by Shen et al. (SH-EOS) \citep{she98a}, which is relatively new, 
is constructed by the relativistic mean field (RMF) theory 
along with a local density approximation.  
The RMF theory is based on the relativistic nuclear many body 
frameworks \citep{bro90} and has been successfully applied 
to the studies of nuclear structures and dense matter \citep{ser86}.  
It is to be noted that the nuclear interactions to derive 
SH-EOS are constrained by the properties of unstable nuclei \citep{sug94}, 
which are available recently in radioactive nuclear beam 
facilities in the world.  
For example, the neutron skin thickness of neutron-rich nuclei, 
which is sensitive to the nuclear symmetry energy, 
is well reproduced for isotopes of light nuclei \citep{tsuz95,sug96}.

The two sets of supernova EOS are different from each other, 
reflecting the characteristic of the nuclear theoretical frameworks and 
the inputs of nuclear data used to constrain the EOS.
The relativistic frameworks tend to provide a stiff EOS 
as compared with the non-relativistic frameworks.  
The density dependence of symmetry energy tends to be 
strong in the relativistic frameworks \citep[See, for example,][]{sum95b}. 
The incompressibility and symmetry energy of SH-EOS 
are 281 MeV and 36.9 MeV, respectively.  
It is to be noted that those values in SH-EOS are not inputs to 
determine the interactions, but the outcome after the 
fitting of nuclear interactions 
to nuclear data of nuclei including neutron-rich ones.  
For LS-EOS, 
we adopt the set of 180 MeV for the incompressibility among three choices.  
The EOS set of this value has been popularly used 
in most of numerical simulations so far.  
The symmetry energy of LS-EOS is 29.3 MeV.  
When those two sets of EOS are adopted for cold neutron stars, 
the maximum masses are 1.8M$_\odot$ and 2.2M$_\odot$ 
for LS-EOS and SH-EOS, respectively, due to the difference 
of stiffness.  
This difference is essential in discussing 
different maximum masses of proto-neutron stars 
in numerical simulations.
Note that the proto-neutron stars contain more leptons 
and are hotter than ordinary neutron stars, and 
therefore, the maximum mass of the former is not the 
same as for the latter.  

As we will demonstrate in $\S$ \ref{sec:EOS-implications}, 
the density and temperature in proto-neutron stars collapsing 
to black holes become extremely high 
beyond the range of EOS originally provided in the table.  
We have extended the EOS table by calculating the necessary 
quantities according to the original formulation of RMF 
for SH-EOS \citep{sum95a}.  
The corresponding EOS table for LS-EOS is prepared 
by using the subroutine provided by Lattimer and Swesty for public use.  
The simple extrapolation of EOS to these extremely high density 
and temperature, where the applicability of the nuclear many 
body frameworks with only the nucleonic degree of freedom is doubtful, 
is apparently oversimplification.
However, since this is the first comparison of different EOS's in 
the rapid 
black hole formation from the massive star, we adopt the 
simplest extension of the currently available sets of EOS.  
There may be the appearance of hyperons, possible condensations 
of mesons and phase transitions to quarks and gluons at these high 
density and temperature.  
They are the targets of upcoming numerical simulations.

\subsubsection{Weak interaction rates}
We adopt the same set of weak interaction rates 
as in the previous study on 15M$_\odot$ star \citep{sum05} 
to facilitate comparisons and numerical checks.  
The basic set of reaction rates follows 
the standard formulation of \citet{bru85}.  
In addition, the plasmon process and the nucleon-nucleon 
bremsstralung are included \citep{sum05}.
The latter reaction has been shown to be an important process 
as a source of $\nu_{\mu/\tau}$ and $\bar{\nu}_{\mu/\tau}$ 
from the proto-neutron star \citep{bur00,suz93}.  
The recent developments of neutrino-matter interactions \citep{bur06} 
and electron-capture rates \citep{lan03} will be implemented 
in future.  
One should be reminded that the details of neutrino signals 
from the black hole formation may be modified by these updated neutrino 
reaction rates and further scrutiny.  
We will demonstrate, however, that 
the whole dynamics is controlled dominantly by the accretion 
and the general feature of neutrino emission reflects largely 
the difference of EOS rather than the minutes of neutrino rates.  

\subsection{Initial Model}
We adopt the presupernova model of 40M$_\odot$ by \citet{woo95}.  
This is the most massive model in the series of their presupernova models 
and contains an iron core of 1.98M$_\odot$.  
Their presupernova model of 15M$_\odot$, which is the {\it standard} 
for recent supernova studies, contains an iron core of 1.32M$_\odot$ 
by contrast.  
The large size of the iron core in the current model warrants 
the black hole formation without explosion as its fate.  
We use the profile of central part of the presupernova model 
up to 3.0M$_\odot$ in baryon mass coordinate, large enough 
to describe the accretion of material for a long time.  

\subsection{Criterion for black hole formation}\label{sec:criterionBH}
The general relativistic formulation for the numerical code 
is based on the metric 
of \citet{mis64}.  
In order to track down the formation of black hole, 
we use the apparent horizon, 
following the numerical studies of black hole formations 
from supermassive stars by \citet{nak06}.  
The apparent horizon exists when the relation 
\begin{equation}
\frac{U}{c} + \Gamma \le 0
\end{equation}
or equivalently 
\begin{equation}
r \le r_g
\end{equation}
is satified, 
where $U$ is the radial fluid velocity and 
$\Gamma$ is the general relativistic gamma factor 
\citep[][for the definitions]{yam97}.
The Schwarzschild radius is defined by
\begin{equation}
r_g = \frac{2 G M_g(r)}{c^2}
\end{equation}
where $M_g(r)$ denotes the gravitational mass inside the radius $r$.
In the current study, we terminate the numerical simulations soon 
after the formation of the apparent horizon because of possible 
numerical errors and instability.  
In order to follow the evolution in a stable manner 
up to the final moment of the fade-out of neutrino emissions, 
we need to apply the singularity-avoiding scheme as in \citet{bau95}.  
We stress that 
the main aim of our study is to clarify the dynamics 
from the beginning of collapse up to the black hole formation 
through the accretion of outer envelopes as well as 
the major part of the neutrino signal during this period 
and, most importantly, their dependence on the EOS.  

\section{Numerical Results}\label{sec:results}
We present the numerical results for the two models 
studied with Shen's EOS and Lattimer-Swesty EOS, 
which are hereafter denoted by SH and LS, respectively.
We start with the outline of the whole evolutions of the two models 
($\S$ \ref{sec:evolution}) 
and discuss each stage from the core collapse, bounce 
($\S$ \ref{sec:bounce}), 
shock propagation, proto-neutron star evolution ($\S$ \ref{sec:pns})
to the formation of black hole ($\S$ \ref{sec:recollapse}).  
We present the results on the distributions and emissions of neutrinos
in $\S$ \ref{sec:nu-distributions} and \ref{sec:neutrino}.

\subsection{Outline of evolution}\label{sec:evolution}
\begin{figure}
\epsscale{.70}
\plotone{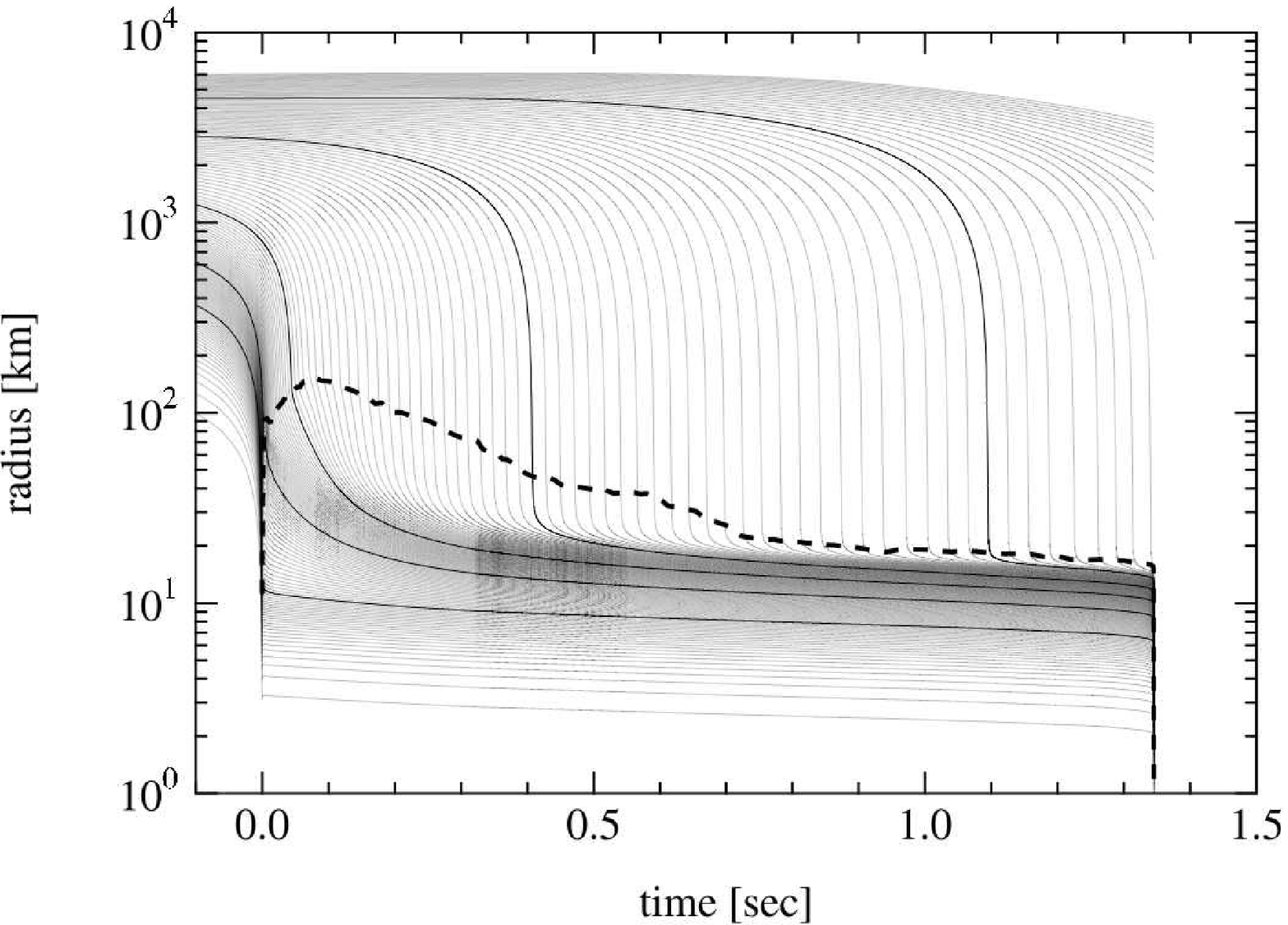}
\caption{Radial trajectories of mass elements 
of the core of 40M$_{\odot}$ star 
as a function of time after bounce in model SH.
The location of shock wave is displayed by a thick dashed line.}
\label{fig:traj-sh}
\end{figure}

\begin{figure}
\epsscale{.70}
\plotone{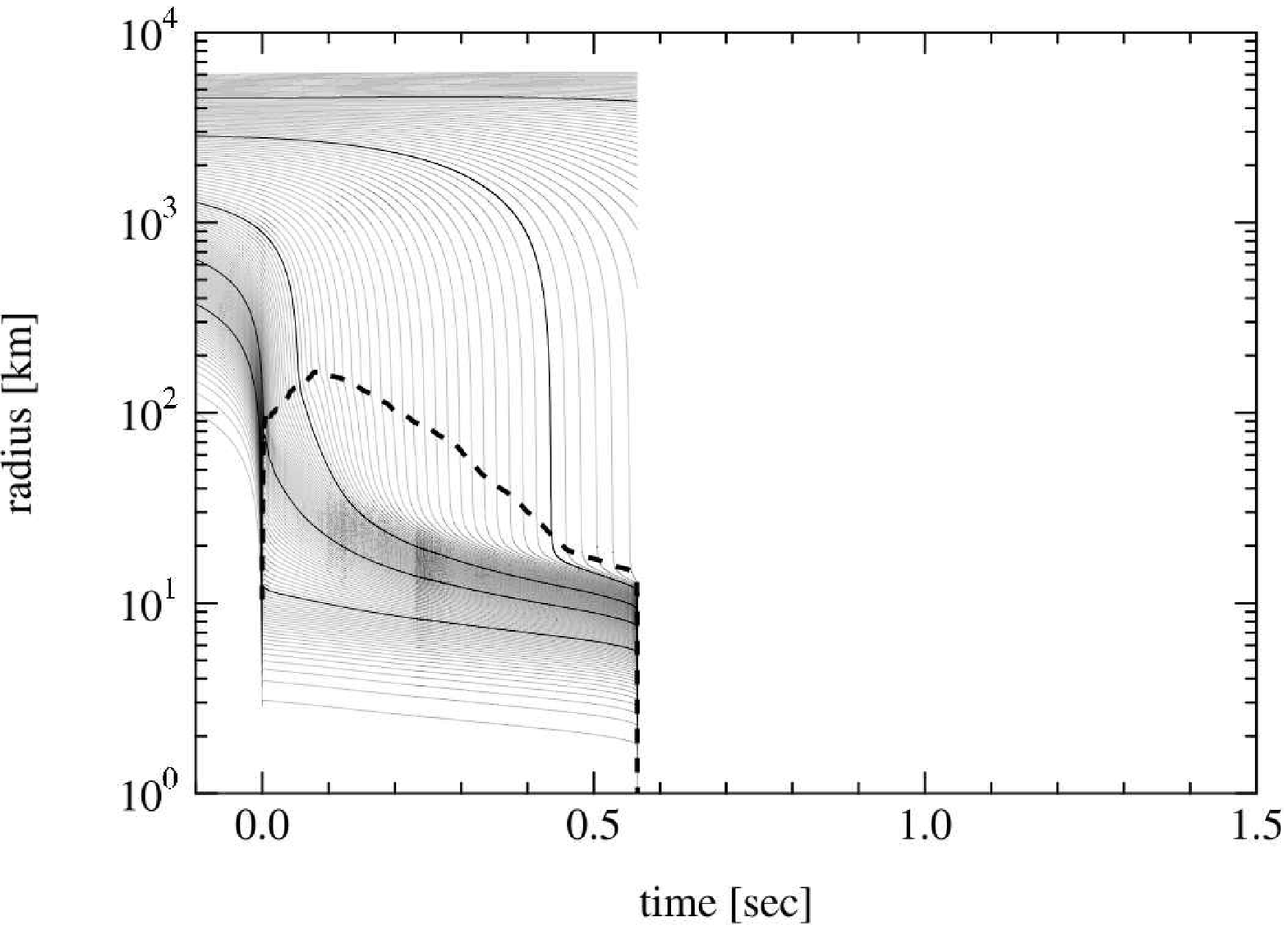}
\caption{Radial trajectories of mass elements 
of the core of 40M$_{\odot}$ star 
as a function of time after bounce in model LS.
The location of shock wave is displayed by a thick dashed line.}
\label{fig:traj-ls}
\end{figure}

Figs. \ref{fig:traj-sh} and \ref{fig:traj-ls} show 
the radial trajectories of mass elements 
as a function of time after core bounce (t$_{pb}$) in models SH and LS.  
The trajectories are plotted for each 0.02M$_{\odot}$ in baryon mass coordinate.  
Thick lines denote the trajectories for 0.5M$_{\odot}$, 1.0M$_{\odot}$, 
1.5M$_{\odot}$, 2.0M$_{\odot}$ and 2.5M$_{\odot}$.  
After the gravitational collapse of central iron core 
of the initial model, 
the core bounce occurs at central densities just above the nuclear matter 
density as in the ordinary core-collapse supernovae.  
The time at the core bounce from the start of 
the simulation is 0.357 sec and 0.333 sec, respectively,
in models SH and LS.  

The shock wave is launched by the core bounce 
and propagates beyond 100 km in both models.
Because of the significant amount of matter accretion, 
the shock wave starts recession at t$_{pb}$ $\sim$100 ms 
and becomes the accretion shock above the central object 
just born at center.
The proto-neutron star is formed after bounce 
and gradually contracts from the beginning.
The whole proto-neutron star shrinks due to the increase 
of mass by the accretion.  
This tendency is striking in model LS, which adopts a softer 
EOS, having large gradients of radial trajectories in the central part.
Note that the rate of accretion is similar in the two models 
as we can see from the trajectories in the outer core above 100 km.  

The gravitational collapse of the proto-neutron star 
occurs when the mass of proto-neutron star exceeds 
the maximum mass for the stable configurations of lepton-rich, 
hot neutron stars.  
The re-collapse proceeds dynamically within a few msec and 
the central object becomes exceedingly compact, leading to 
the formation of black hole.  
The dynamical collapse to the black hole occurs 
at t$_{pb}$=1.34 s and 0.56 s, 
respectively, in models SH and LS.  

\subsection{Initial collapse and bounce}\label{sec:bounce}
The gravitational collapse of central iron core from the initial model 
occurs in a similar way as in the ordinary core-collapse 
supernovae (e.g. 15M$_{\odot}$ model).  
The electron captures on free protons and nuclei 
proceed to reduce the lepton fraction until neutrino trapping.  
The neutrinos ($\nu_e$) are trapped 
at high densities ($\sim$10$^{12}$ g/cm$^3$) 
and contribute to build up the bounce core.  
During the collapse, the compositional difference 
between the two models can be seen 
as we have found in the case of 15M$_{\odot}$ model.
In fact, model SH tends to reduce electron captures, 
having a smaller free proton fraction than model LS.

We define the core bounce (t$_{pb}$=0 s) as the time 
when the central density reaches the maximum in this stage.
The central density at bounce is 3.2$\times$10$^{14}$ g/cm$^3$ 
and 4.1$\times$10$^{14}$ g/cm$^3$ in models SH and LS, respectively.  
After a slight expansion at bounce, the central densities 
gradually increase owing to the contraction of proto-neutron stars.

\begin{figure}
\epsscale{.65}
\plotone{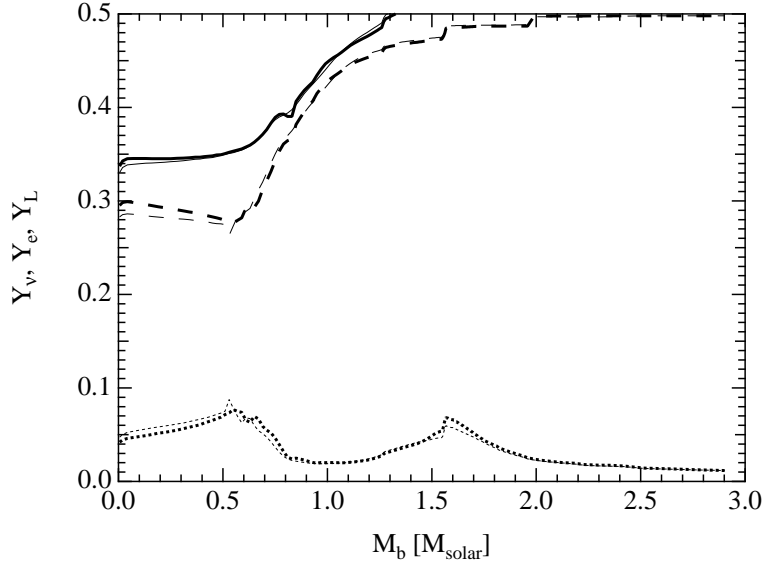}
\caption{Lepton, electron and neutrino ($\nu_e$) fractions at bounce are shown 
as a function of baryon mass coordinate by solid, dashed 
and dotted lines, respectively.
The results for models SH and LS are shown by thick and thin lines, 
respectively.}
\label{fig:tpb0lepton}
\end{figure}

\begin{figure}
\epsscale{.65}
\plotone{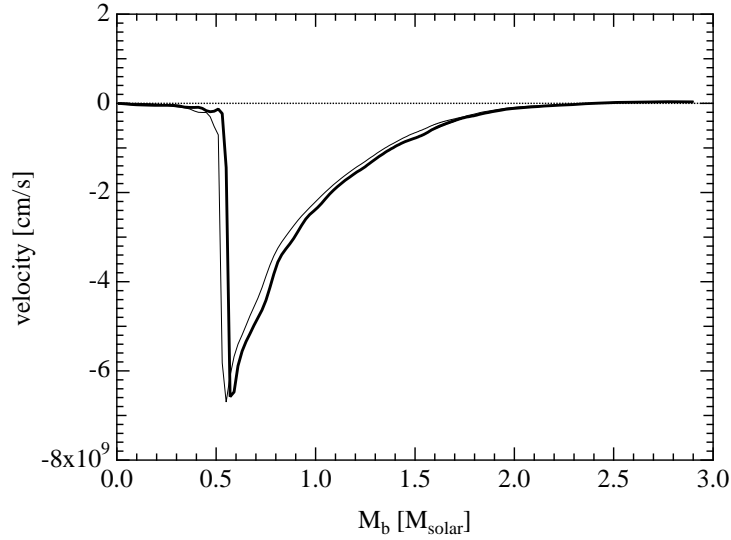}
\caption{Velocity profiles at bounce are shown as a function 
of baryon mass coordinate.
The results for models SH and LS are shown by thick and thin lines, 
respectively.}
\label{fig:tpb0velocity}
\end{figure}

The profiles of lepton fractions at bounce are shown in 
Fig. \ref{fig:tpb0lepton}.  
The lepton fraction and electron fraction at center in SH 
is 0.35 and 0.30, respectively.  
The corresponding values in LS are smaller, 
but the difference between the two models is $\sim$0.01.  
The modest difference leads to a small difference of the bounce cores.  
Fig. \ref{fig:tpb0velocity} displays the velocity profiles 
at bounce in the two models.
The size of bounce core in SH is slightly larger 
than that in LS, but the difference is again small $\le$0.02M$_{\odot}$.  
The radial position of shock wave in SH is 11 km and 
that in LS is only $\sim2 \%$ smaller at their formations.
Accordingly, the propagations of shock wave in the early phase 
(up to t$_{pb}$ $\sim$100 ms) 
are similar to each other in the two models.  

\subsection{Rapid formation of proto-neutron star}\label{sec:pns}
The shock wave propagates through the iron core, 
but it never overcomes the ram pressure of falling material.  
It stalls at $\sim$150 km and turns into the recession 
due to the substantial influence of the accretion.  
In fact, 
the material of about 1M$_{\odot}$ falls down 
within 100 ms after bounce in both models.
The shock wave reaches 1.6M$_{\odot}$ at t$_{pb}$=100 ms 
and the central core below the shock wave acquires 
a typical neutron star mass already.  
Although the outer part of the central core is not yet 
hydrostatic, 
the proto-neutron star has been formed already 
in terms of mass.  
This is much faster than the case of ordinary core-collapse 
supernovae, in which it takes 300--400 ms to acquire 
the enough mass ($\sim$1.5M$_{\odot}$ in baryon mass) 
for proto-neutron stars.

\begin{figure}
\epsscale{.65}
\plotone{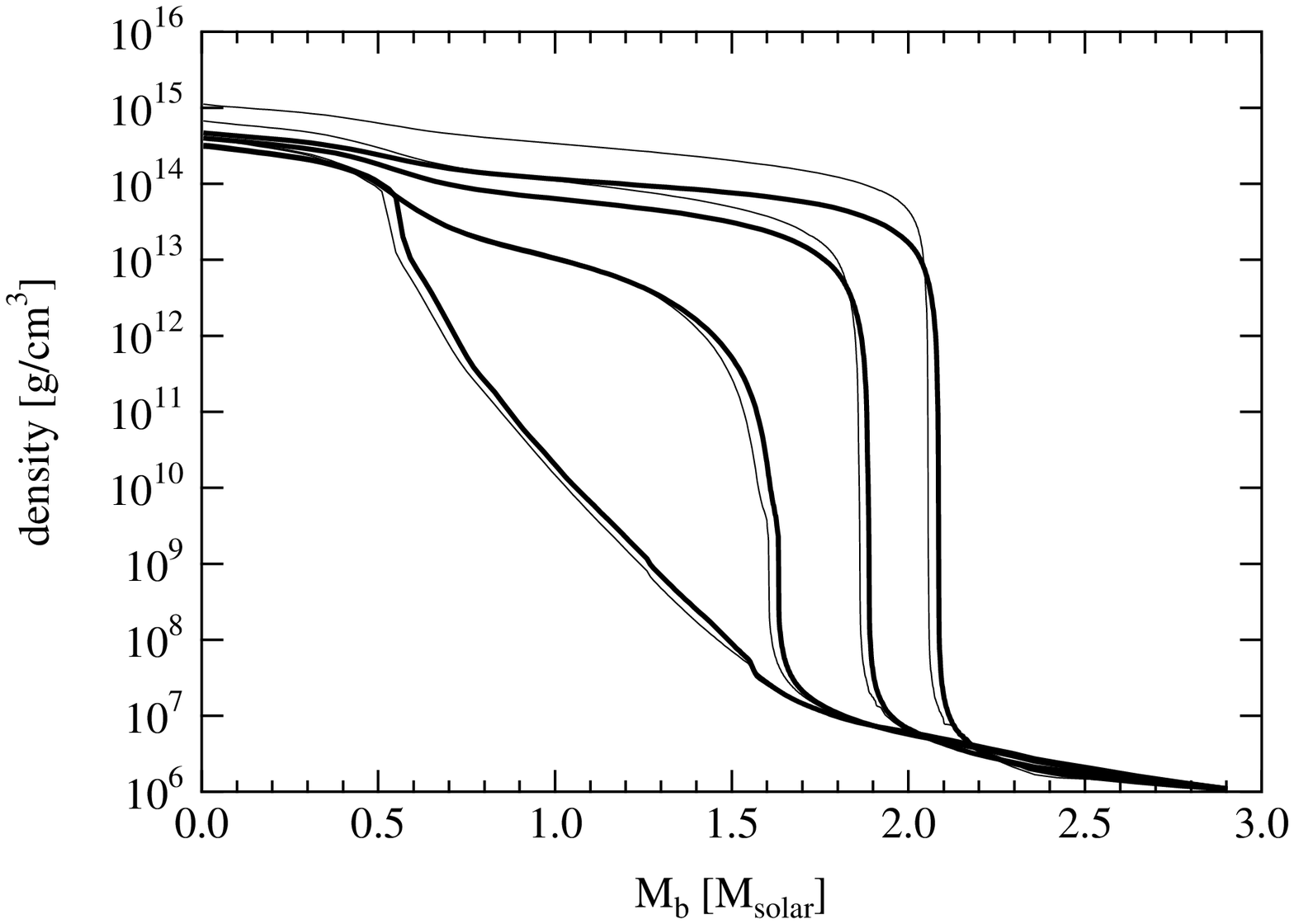}
\caption{Density profiles at t$_{pb}$=0, 100, 300 and 500 ms 
are shown as a function of baryon mass coordinate.
The results for models SH and LS are shown by thick and thin lines, 
respectively.}
\label{fig:tpb0-500density}
\end{figure}

\begin{figure}
\epsscale{.65}
\plotone{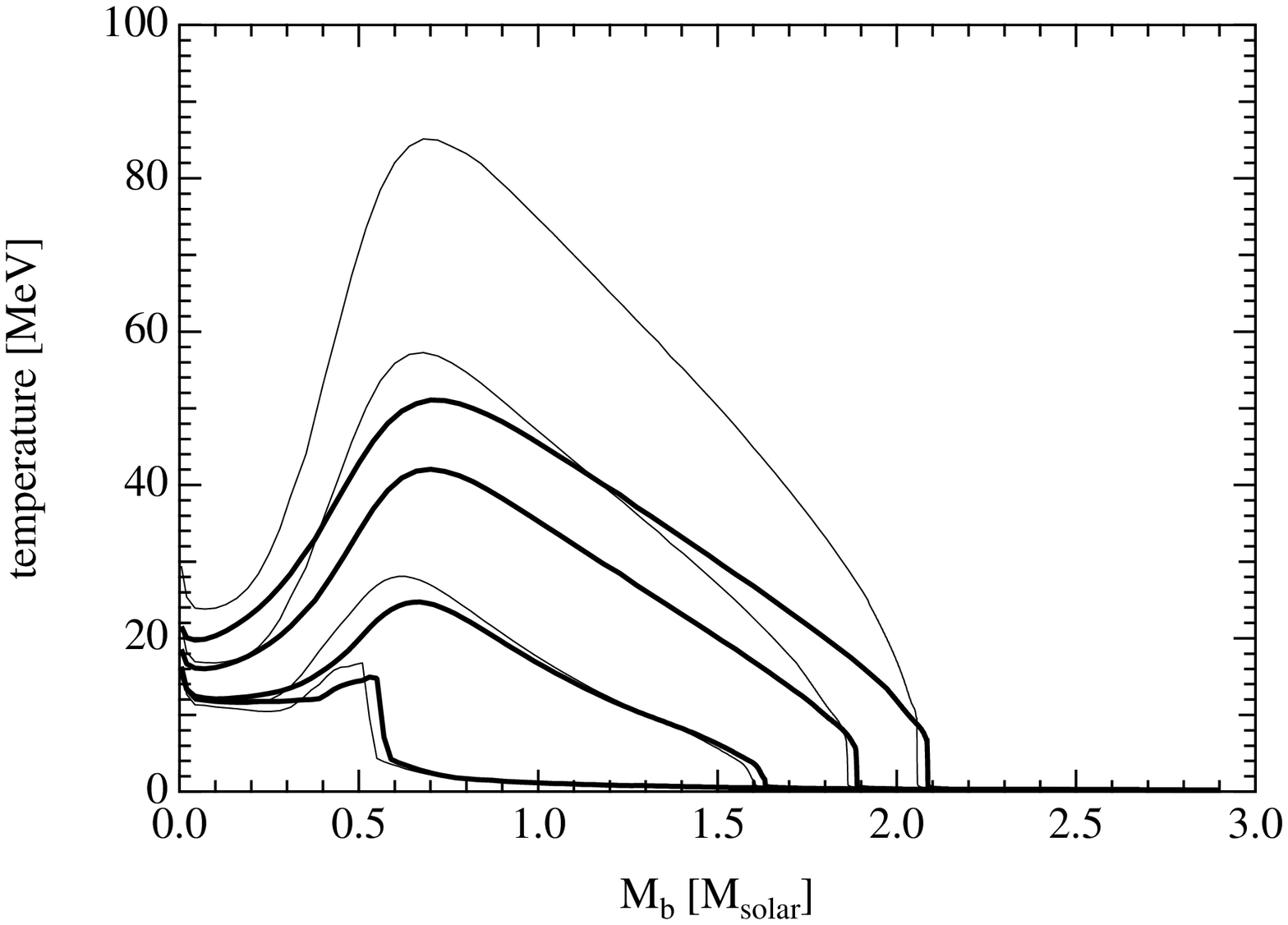}
\caption{Temperature profiles at t$_{pb}$=0, 100, 300 and 500 ms 
are shown as a function of baryon mass coordinate.
The results for models SH and LS are shown by thick and thin lines, 
respectively.}
\label{fig:tpb0-500temperature}
\end{figure}

We show in Figs. \ref{fig:tpb0-500density} and \ref{fig:tpb0-500temperature} 
the profiles of density and temperature in the two models 
at t$_{pb}$=0, 100, 300 and 500 ms.
The rapid growth of proto-neutron star masses is evident 
with the mass over 2M$_{\odot}$ in 500 ms after bounce.
The difference of masses in the two models are within 0.03M$_{\odot}$, 
reflecting the initial difference in the size of bounce cores and 
the similarity of the accretion rates in the two models.
The densities inside proto-neutron star in model LS after t$_{pb}$=300 ms 
become higher than those in model SH 
because of the softness of LS-EOS.  
The central density in model LS exceeds 10$^{15}$ g/cm$^{3}$ at t$_{pb}$=500 ms.
Owing to the compression of dense matter, the temperature inside becomes 
high, peaked at 0.6--0.7M$_{\odot}$, where the heating 
due to the passage of shock wave is significant.  
The peak temperatures reach 51 MeV and 85 MeV at t$_{pb}$=500 ms 
in models SH and LS, respectively.  
This large difference is caused by the faster compression in model LS 
than in model SH.  
We remark that the peak values of entropy per baryon in the two models 
are almost equal.  

\begin{figure}
\epsscale{.65}
\plotone{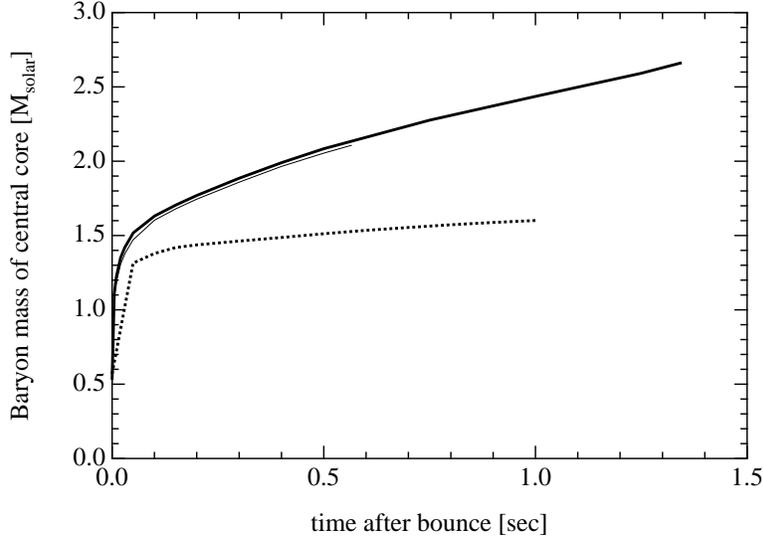}
\caption{Baryon masses of central core (proto-neutron star) 
in model SH (thick solid line) and LS (thin solid line) 
are shown as a function of time after bounce.
The case of a 15M$_{\odot}$ star with SH-EOS is plotted 
by dotted line for comparison.}
\label{fig:mass-core}
\end{figure}

The time evolutions of the baryon mass inside the proto-neutron star 
in the two models 
are shown in Fig. \ref{fig:mass-core}.
They are surprisingly similar 
after bounce, which means that the accretion rates are 
similar in the two models.
In model LS, the baryon mass reaches 2.10M$_{\odot}$ at t$_{pb}$=0.56 s 
and exceeds the maximum mass for the stable configurations, 
leading to the dynamical collapse to the black hole.
In model SH, it increases further up to 2.66M$_{\odot}$ at t$_{pb}$=1.34 s.
For comparison, we plot the corresponding baryon mass 
of the proto-neutron star taken from the numerical result of 
the gravitational collapse of a 15M$_{\odot}$ star 
with SH-EOS \citep{sum05} as an example of massive stars,
which are supposed to lead to ordinary core-collapse supernovae.
The growth of proto-neutron star is much faster in the case 
of the 40M$_{\odot}$ star than in the case of the 15M$_{\odot}$ star.  
This means that the accretion rate in the current models of 40M$_{\odot}$ 
is considerably higher than that in the canonical model of 15M$_{\odot}$.  
The accretion rate is $\sim$1M$_{\odot}$/s in the current 
models at t$_{pb}$=0.4 s 
while it is $\sim$0.2M$_{\odot}$/s for 15M$_{\odot}$ model.
This large accretion rate makes the astrophysical events 
from the 40M$_{\odot}$ stars unique, 
having a rapid contraction of proto-neutron stars 
and a quick formation of black hole.  
This is clearly different from the delayed scenario of black hole formation 
after the cooling of proto-neutron stars from ordinary supernovae.

\subsection{Re-collapse and formation of black hole}\label{sec:recollapse}
\begin{figure}
\epsscale{1.1}
\plottwo{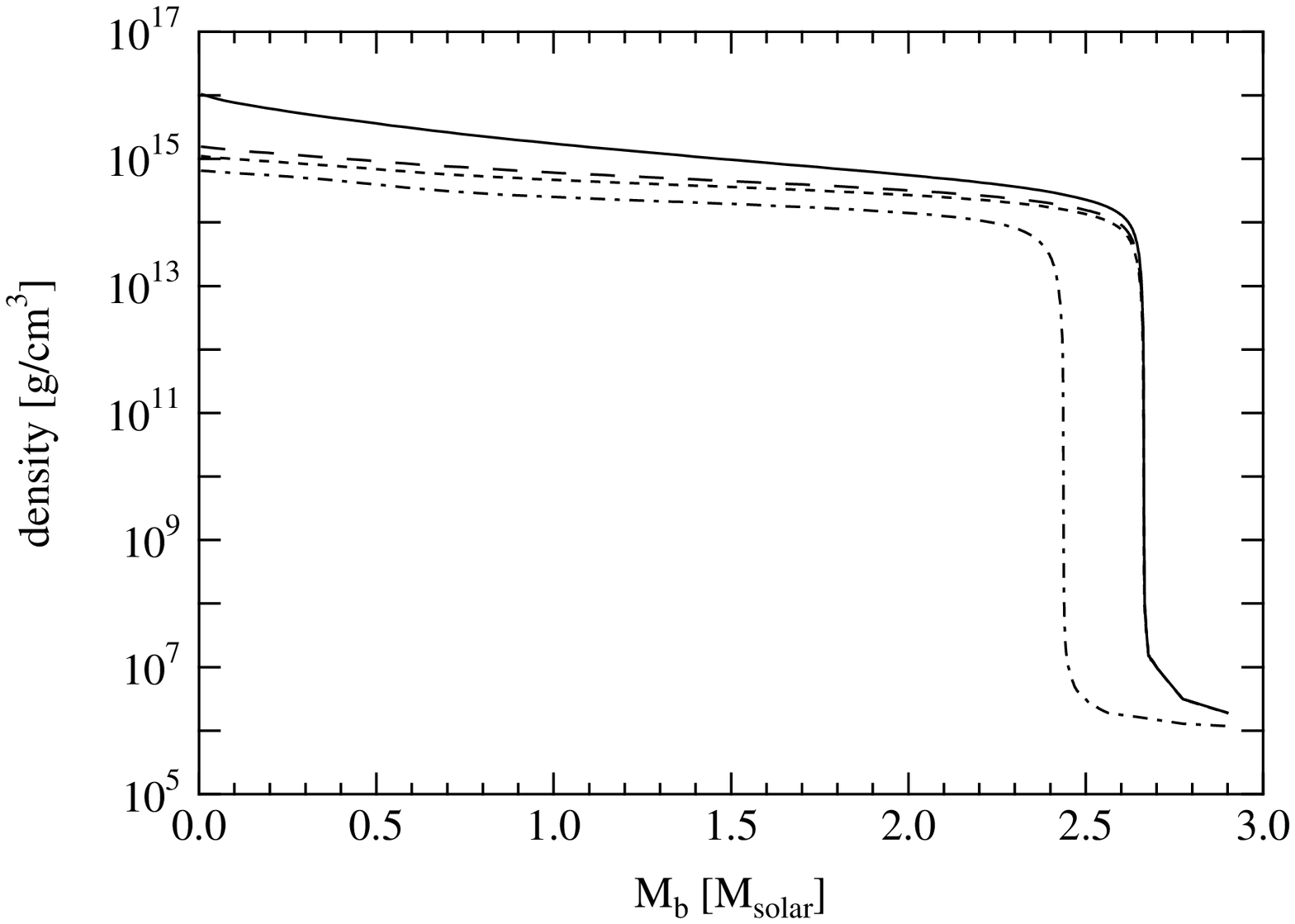}{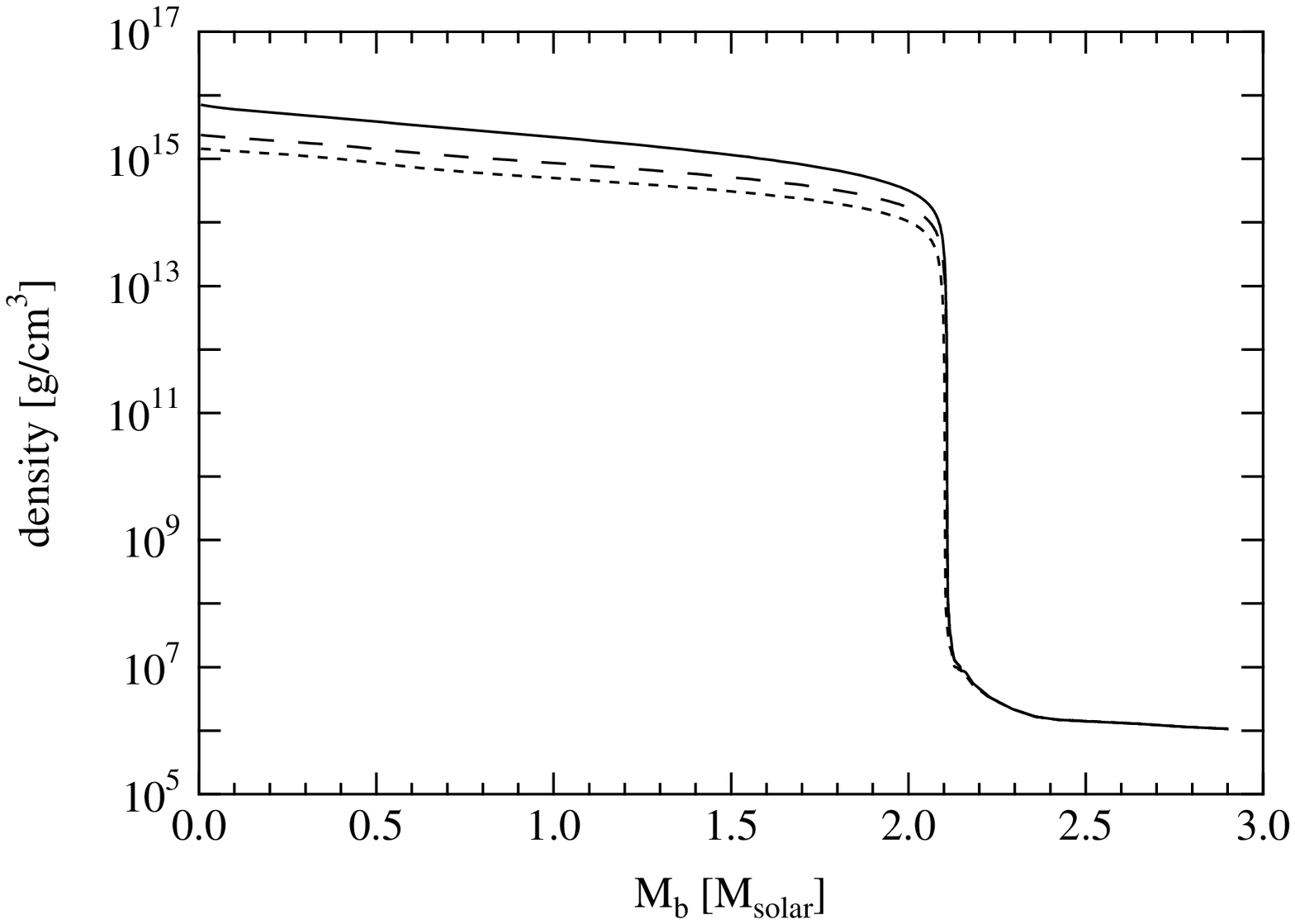}
\caption{Density profiles are shown as a function 
of baryon mass coordinate in models SH (left) and LS (right).
The profiles at the formation of apparent horizon are depicted by solid lines.
The profiles at the beginning of re-collapse and 
at the time when the central 
density reaches $\sim$2$\times$10$^{15}$ g/cm$^3$ 
are depicted by dashed and long-dashed lines, respectively.
For model SH, the profile at t$_{pb}$=1 s is shown by dot-dashed line.}
\label{fig:bh-density}
\end{figure}

\begin{figure}
\epsscale{1.1}
\plottwo{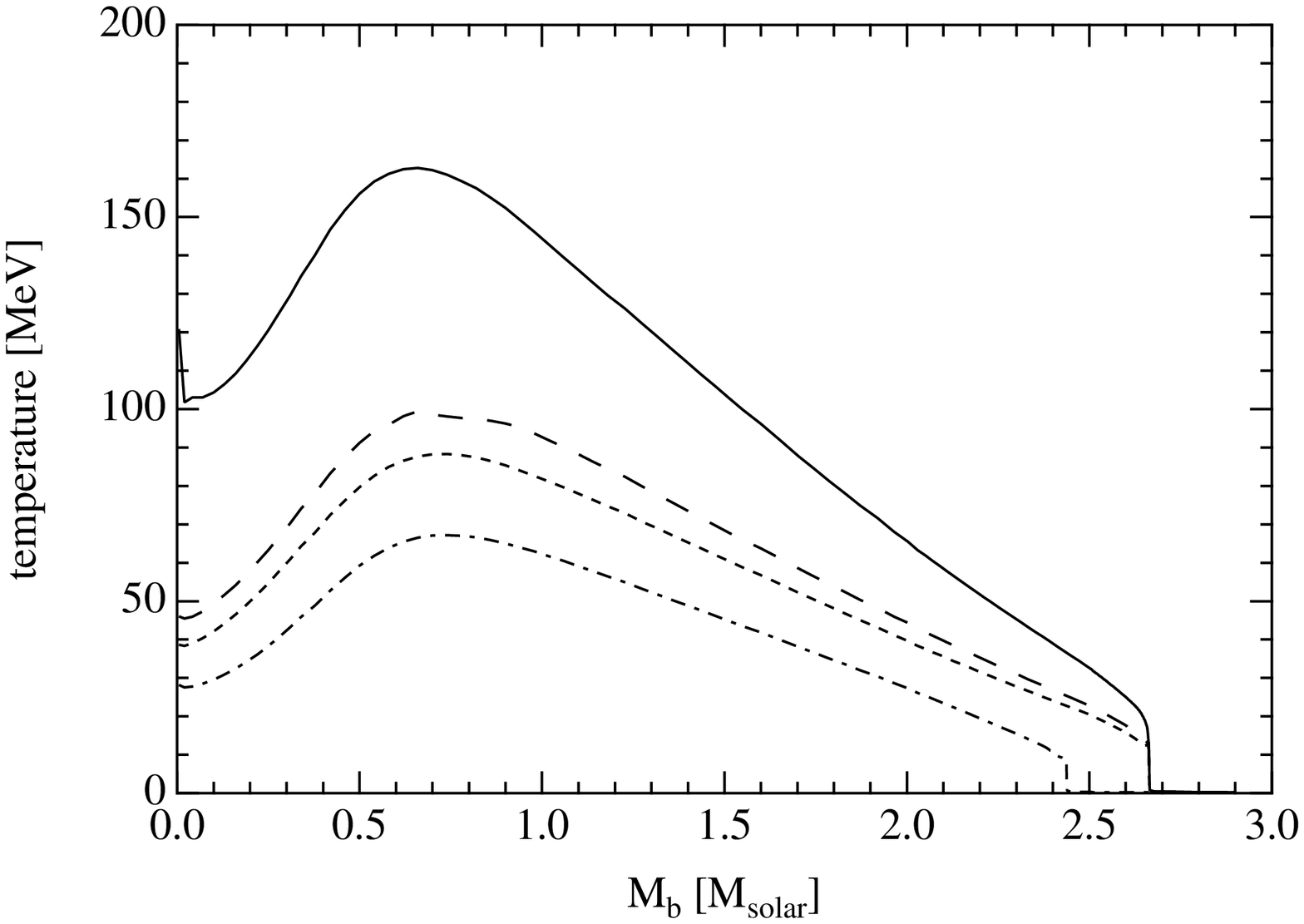}{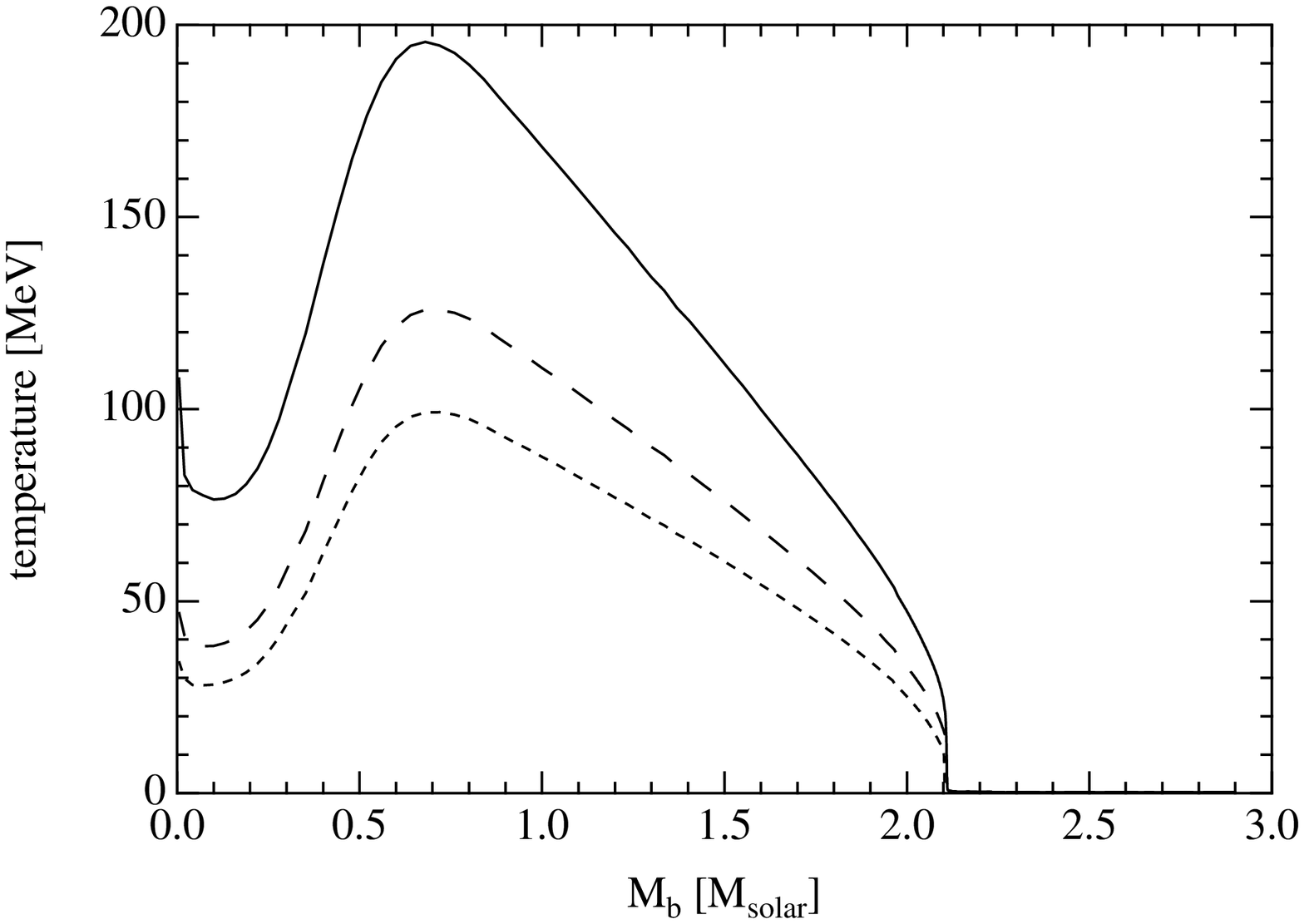}
\caption{Temperature profiles are shown as a function 
of baryon mass coordinate in models SH (left) and LS (right).
The notation is the same as in Fig. \ref{fig:bh-density}.}
\label{fig:bh-temperature}
\end{figure}

\begin{figure}
\epsscale{1.1}
\plottwo{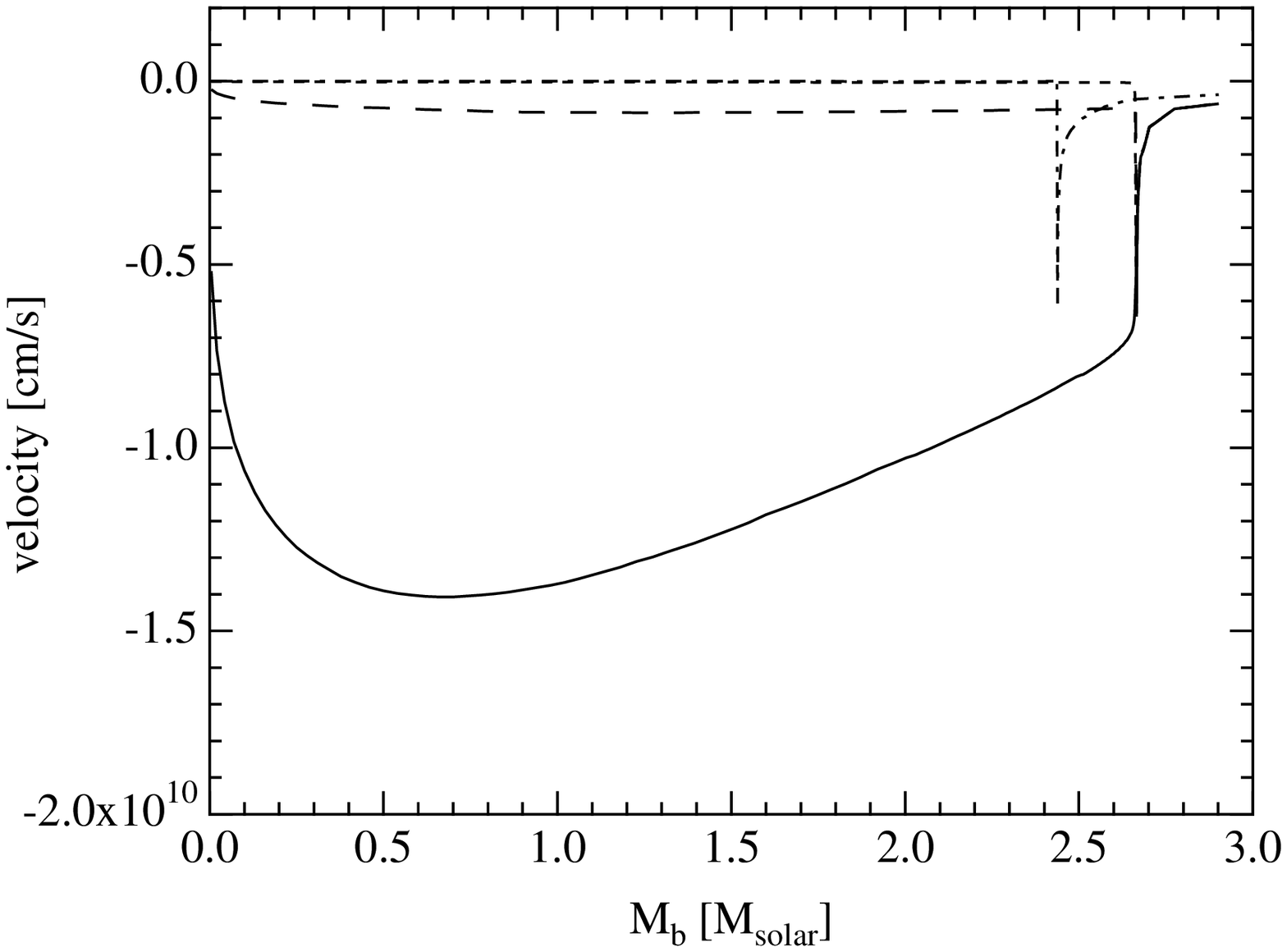}{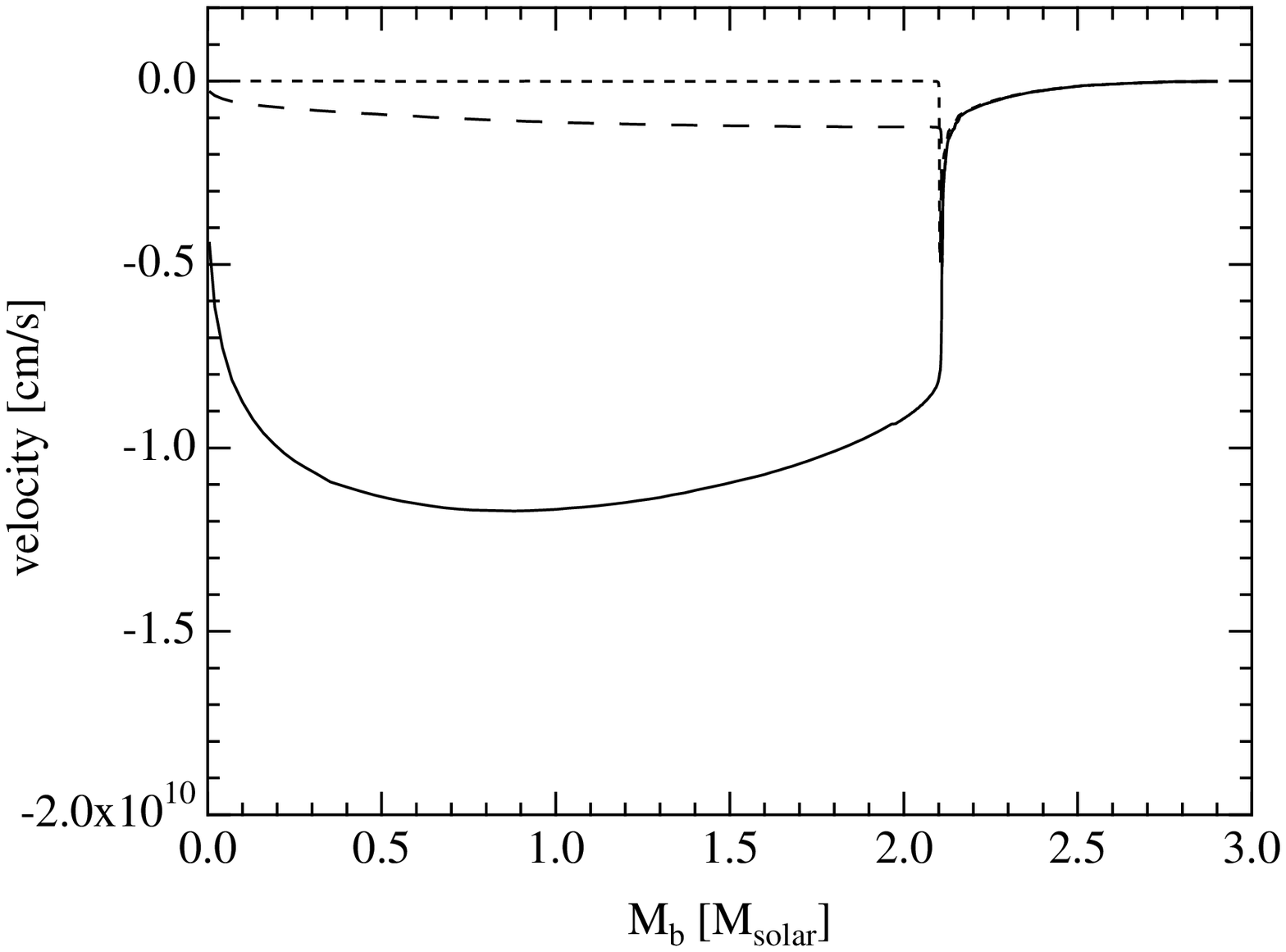}
\caption{Velocity profiles are shown as a function 
of baryon mass coordinate in models SH (left) and LS (right).
The notation is the same as in Fig. \ref{fig:bh-density}.}
\label{fig:bh-velocity}
\end{figure}

In Figs. \ref{fig:bh-density},  \ref{fig:bh-temperature} and  
\ref{fig:bh-velocity}, we display the profiles of density, temperature 
and velocity at final stages up to the black hole formation.  
Here t$_{bh}$ denotes the time before the black hole formation.  
As sample snapshots, 
we plot the profiles at the beginning of re-collapse 
(t$_{bh}$=-8.2 ms and -1.0 ms for SH and LS, respectively) 
by dashed lines 
and the profiles at the time 
(t$_{bh}$=-0.14 ms and -0.19 ms for SH and LS, respectively) 
when the central density reaches 
$\sim$2$\times$10$^{15}$ g/cm$^3$ (near the maximum density in the 
original table of SH-EOS) 
by long-dashed lines.  
The timing of the snapshot of the re-collapse is chosen to be 
when the velocity just behind the shock wave exceeds $-2\times10^{8}$ cm/s 
for practical reasons.
The profiles at the formation of apparent horizon are shown 
by solid lines.  
For model SH, we plot also the profile at t$_{pb}$=1 s by dot-dashed lines.

In both models, the density in the whole proto-neutron stars increases 
within short times during the re-collapse, 
while the baryon mass of proto-neutron stars does not change so much.
The central density reaches $\sim$10$^{16}$ g/cm$^{3}$ at the end 
of computations.  
The temperature increases beyond 100 MeV during the re-collapse and 
reaches $\sim$200 MeV at the peak position for model LS and 
a lower value for model SH.  
The re-collapse proceeds dramatically having the velocity increase 
from $\sim 10^{6}-10^{8}$ cm/s at the beginning, 
to $\sim$10$^{9}$ cm/s within $1-10$ ms.  
The velocity eventually exceeds $\sim$10$^{10}$ cm/s 
by the time of black hole formation.  

\begin{figure}
\epsscale{1.1}
\plottwo{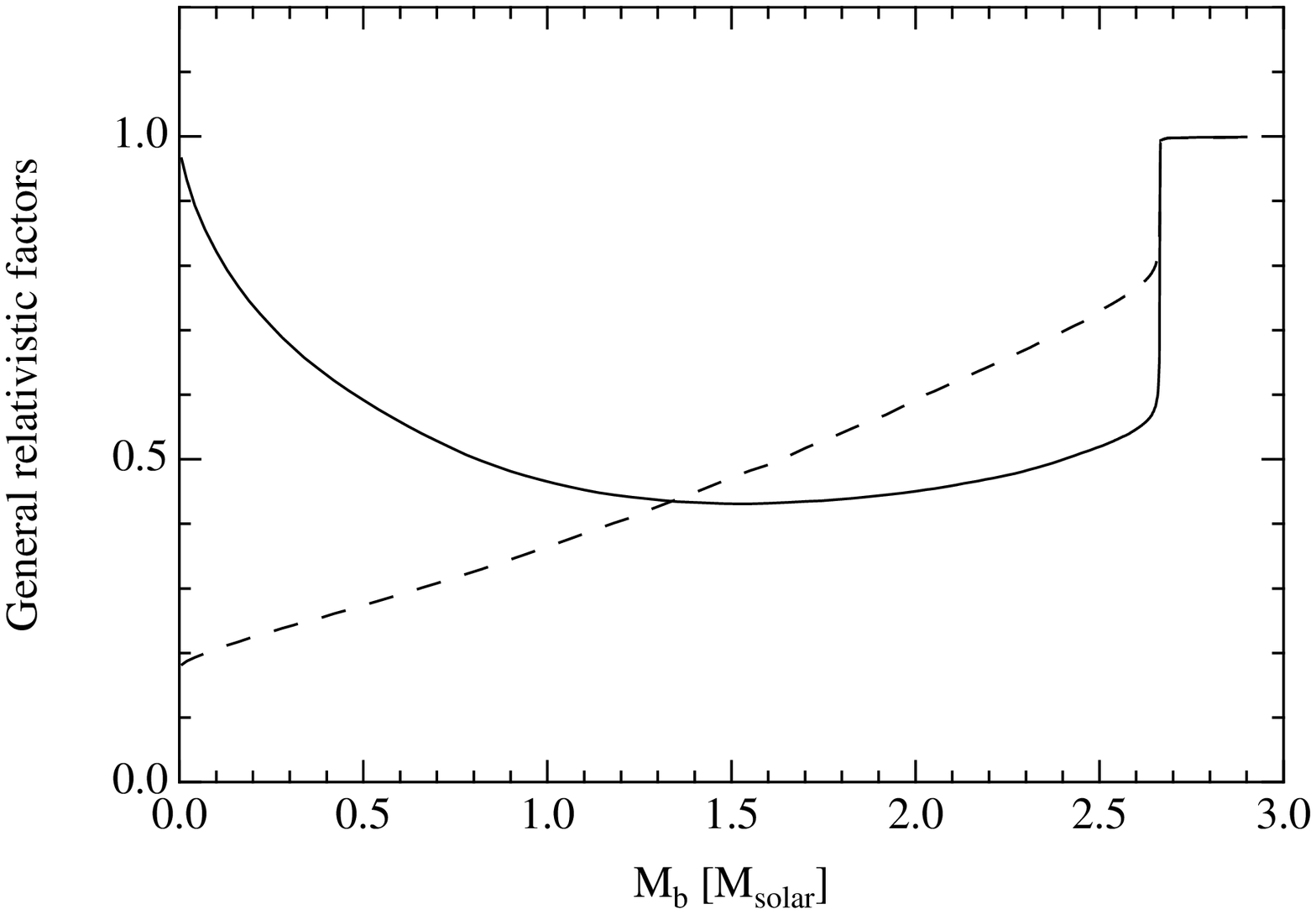}{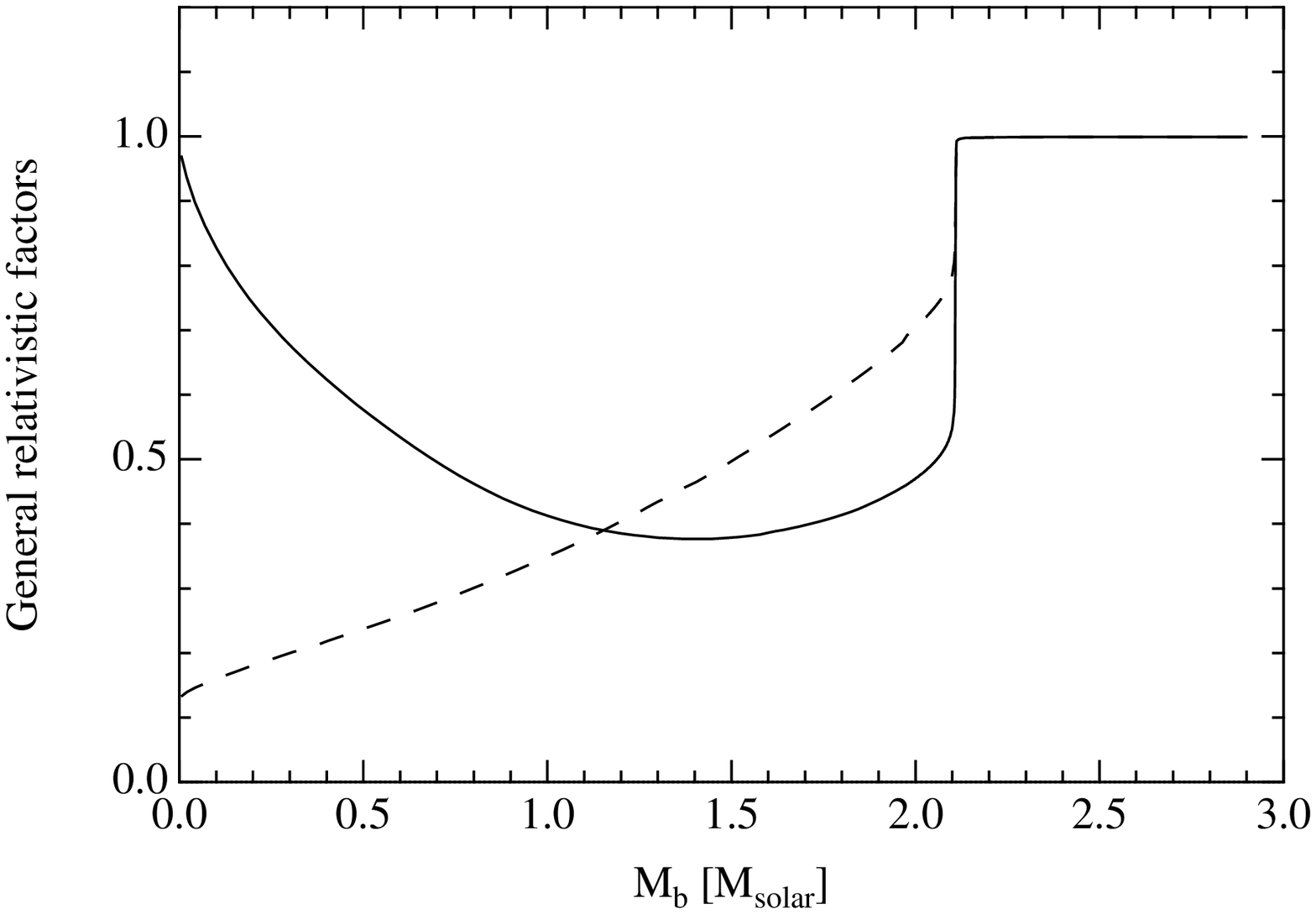}
\caption{General relativistic factors, $\Gamma$ and $e^{\phi}$, 
at the formation of apparent horizon
are shown as a function of baryon mass coordinate 
by solid and dashed lines, respectively, 
in models SH (left) and LS (right).}
\label{fig:bh-grfactor}
\end{figure}

The general relativistic factors, $\Gamma$ and $e^{\phi}$, at the time 
of formation of apparent horizon are shown in Fig. \ref{fig:bh-grfactor}.  
The general relativistic gamma factor, $\Gamma$, has the minimum 
around 1.5M$_{\odot}$.
The large negative values for the velocity affects the location of 
apparent horizon, which is defined by the condition, $U/c + \Gamma  = 0$ 
(see the definition in $\S$ \ref{sec:criterionBH}).  
The apparent horizon is formed 
at $\sim$1.2M$_{\odot}$ ($\sim$5 km in radius) in model SH and 
at $\sim$1.3M$_{\odot}$ ($\sim$5 km in radius) in model LS.  

\subsection{Neutrino distributions}\label{sec:nu-distributions}
\begin{figure}
\epsscale{0.6}
\plotone{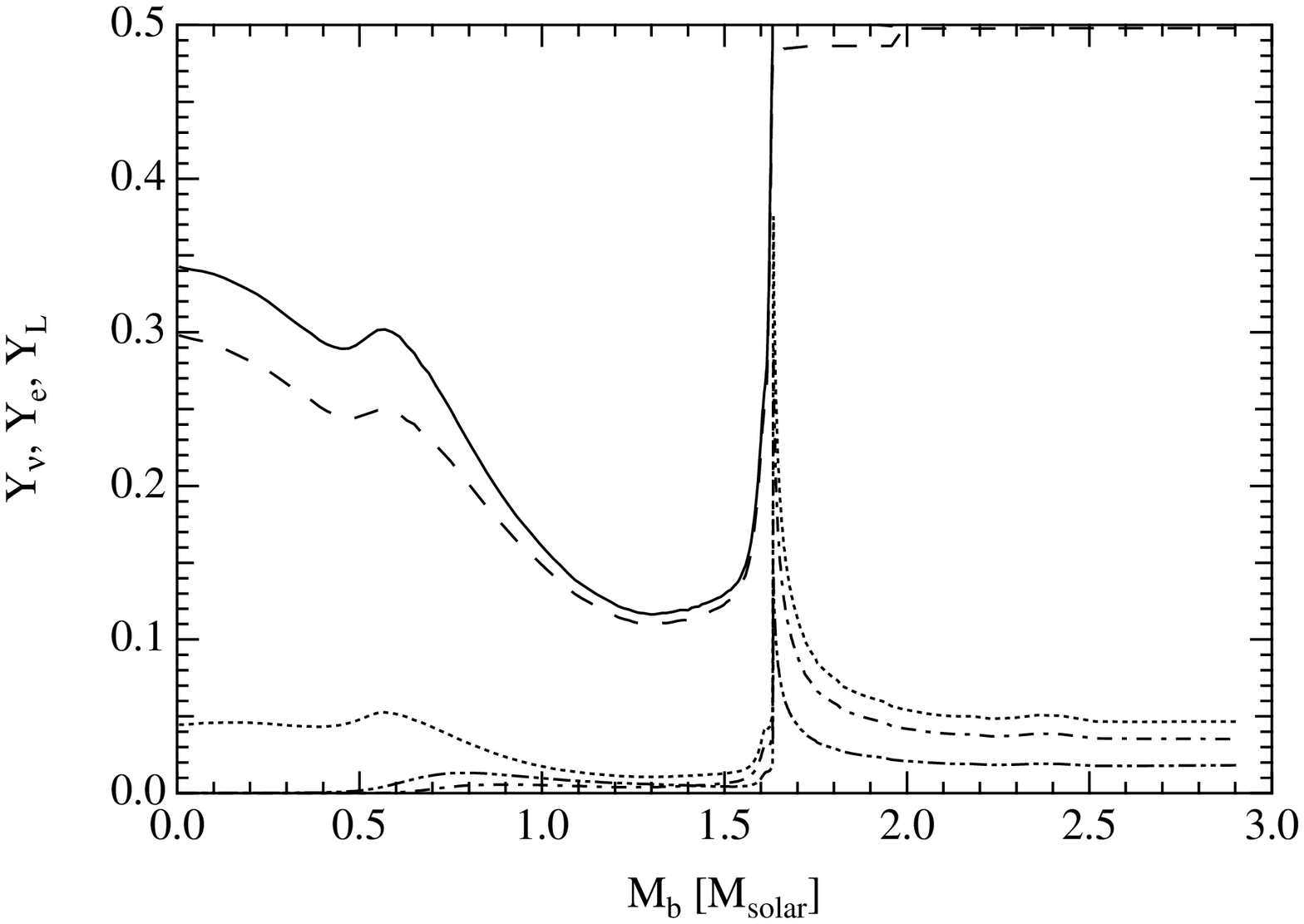}
\plotone{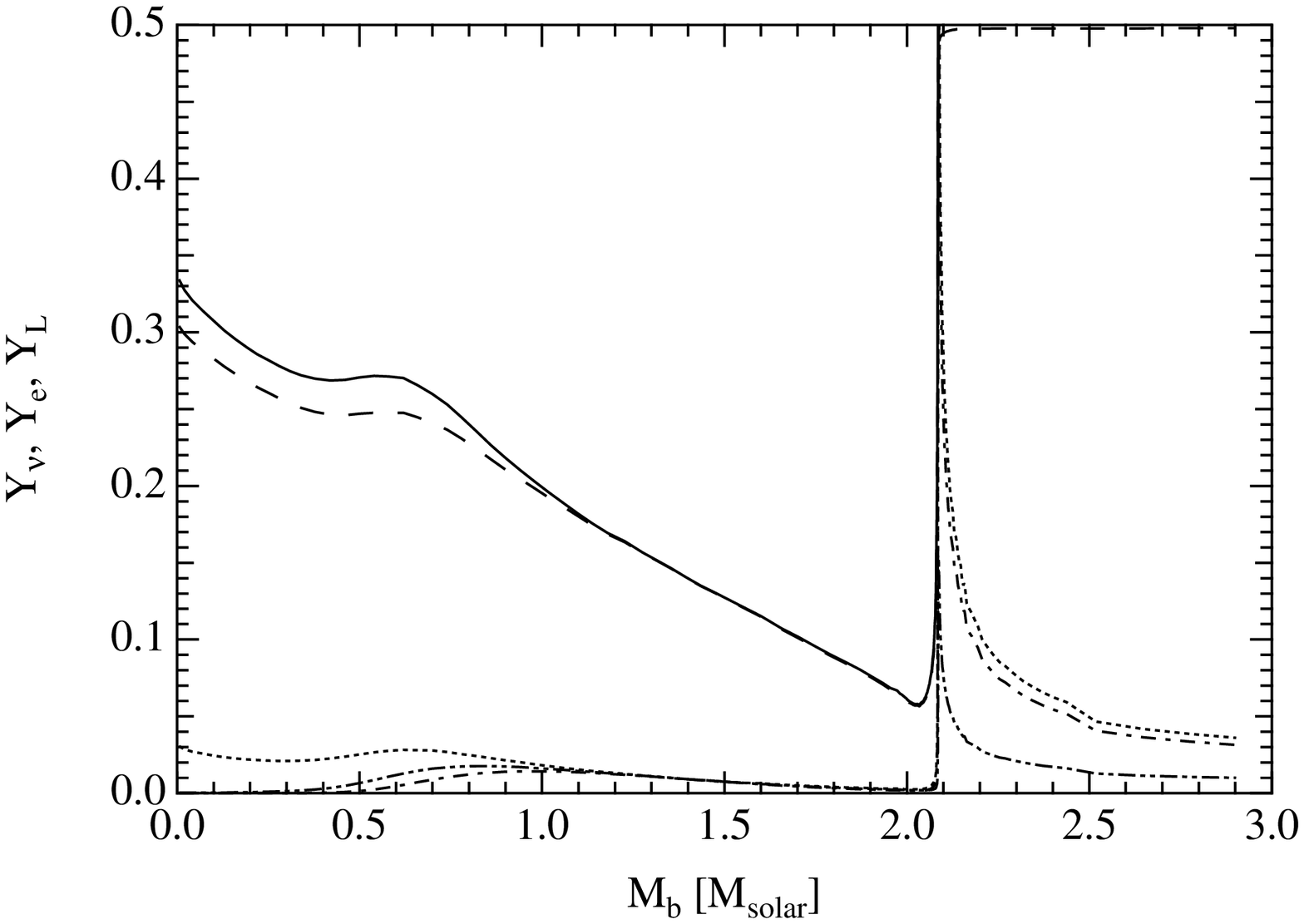}
\plotone{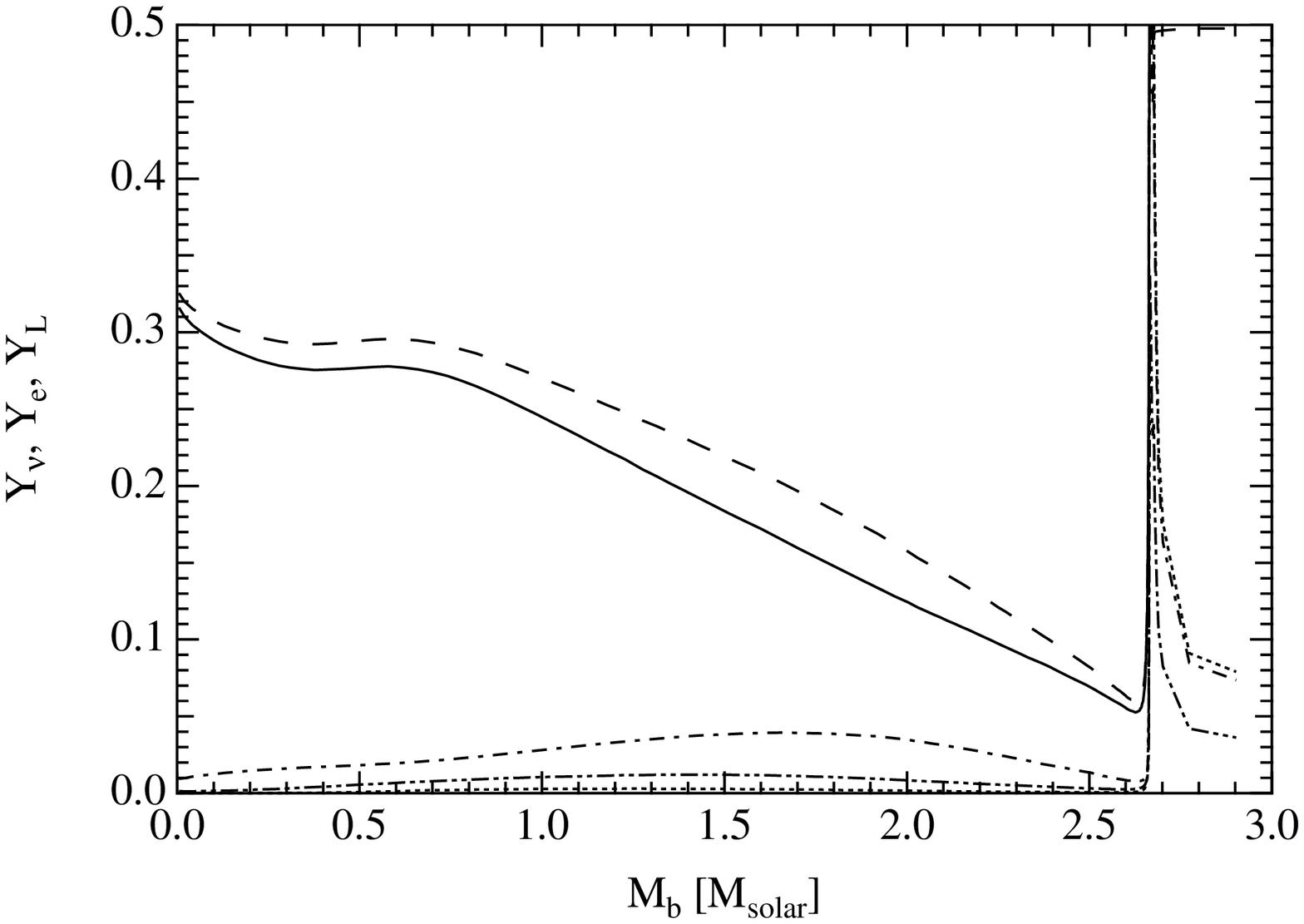}
\caption{Number fractions of leptons, electrons, 
$\nu_e$, $\bar{\nu}_e$ and $\nu_{\mu/\tau}$ 
are shown as a function of baryon mass coordinate 
by solid, dashed, dotted, dot-dashed and dot-dot-dashed lines, respectively, 
at t$_{pb}$=100 msec (top), t$_{pb}$=500 msec (middle) 
and t$_{bh}$=0 (bottom) for model SH.}
\label{fig:sh-ylepton}
\end{figure}

We show the distributions of leptons in model SH 
at t$_{pb}$=100, 500 msec and t$_{bh}$=0 (black hole formation) 
in Fig. \ref{fig:sh-ylepton}.  
The number fraction (number per baryon), Y$_{i}$, is defined 
by the ratio, n$_{i}$/n$_{B}$, 
where n$_{i}$ and n$_{B}$ denote the number densities of 
particle $i$ and baryon, respectively.
The (electron-type) lepton fraction, Y$_{L}$, which is defined by the sum of 
net electron fractions, Y$_{e^{-}}-$Y$_{e^{+}}$, and 
net (electron-type) neutrino fractions, Y$_{\nu_{e}}-$Y$_{\bar\nu_{e}}$, 
is shown by solid line.  
The (net) electron fraction, Y$_{e}$, neutrino fractions for 
$\nu_e$, $\bar{\nu}_e$ and $\nu_{\mu/\tau}$ are shown 
by dashed, dotted, dot-dashed and dot-dot-dashed lines, respectively.

After the core bounce, the neutronization of matter proceeds 
after the passage of shock wave and the electron fraction 
decreases drastically like in the case of the proto-neutron star formation 
in ordinary supernovae.  
A trough of the lepton fraction is formed 
between the outer edge of inner core and the shock position 
by t$_{pb}$=100 msec and persists until the black hole formation.  
The profile of the lepton fraction is not changed much 
while the shape becomes wider according to the shock propagation.  
The fraction of $\nu_e$, which has been produced during the collapse 
and trapped inside the inner core, decreases during the evolution.  
The $\bar{\nu}_e$ fraction increases initially in the outer core 
and prevails in the whole core at the end.  
Because of the appearance of $\bar{\nu}_e$'s, the lepton fraction 
is smaller than the electron fraction at t$_{bh}$=0.  
The $\nu_e$ and $\bar{\nu}_e$ fractions are determined 
by the conditions for beta equilibrium including neutrinos and 
their evolution is driven 
by the decrease of chemical potential, $\mu_{\nu}$, 
for electron-type neutrinos.
Since the relation among the chemical potentials for electrons, 
protons and neutrons holds as 
$\mu_{\nu}=\mu_{e}+\mu_{p}-\mu_{n}$ 
through the quasi-chemical equilibrium, 
the reduction of $\mu_{\nu}$ occurs by the compression of matter 
leading to the black hole formation.  
This is because the compression leads to the increase of $\mu_{n}$ 
and the decrease of $\mu_{p}$ by the increasing effect of 
symmetry energy at very high densities.  
The decrease of $\mu_{\nu}$ (actually to negative values) 
drives a shift from the dominance 
of $\nu_e$ to that of $\bar{\nu}_e$ under the beta equilibrium.
The $\nu_{\mu/\tau}$ and $\bar \nu_{\mu/\tau}$ neutrinos are 
produced by pair processes and, therefore, appear mainly 
in the region with non-degenerate electrons.  

\begin{figure}
\epsscale{0.6}
\plotone{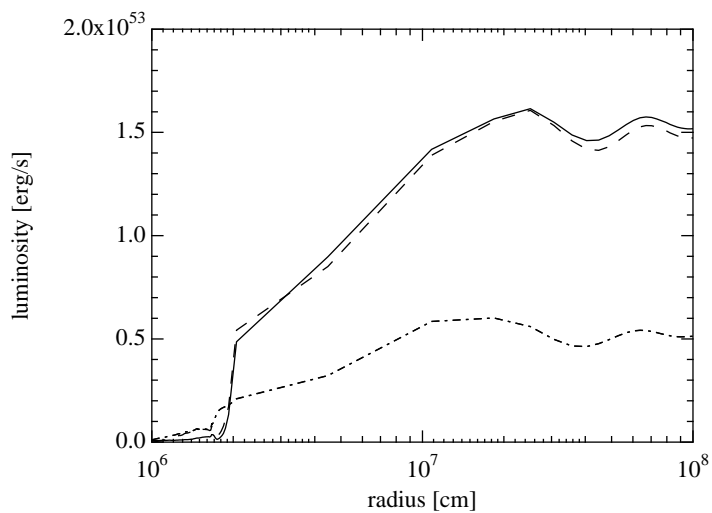}
\caption{Luminosities of 
$\nu_e$, $\bar{\nu}_e$ and $\nu_{\mu/\tau}$ 
are shown as a function of radius 
by solid, dashed and dot-dashed lines, respectively, 
at t$_{pb}$=1 sec for model SH.}
\label{fig:sh-tpb1000-lumi}
\end{figure}

\begin{figure}
\epsscale{0.6}
\plotone{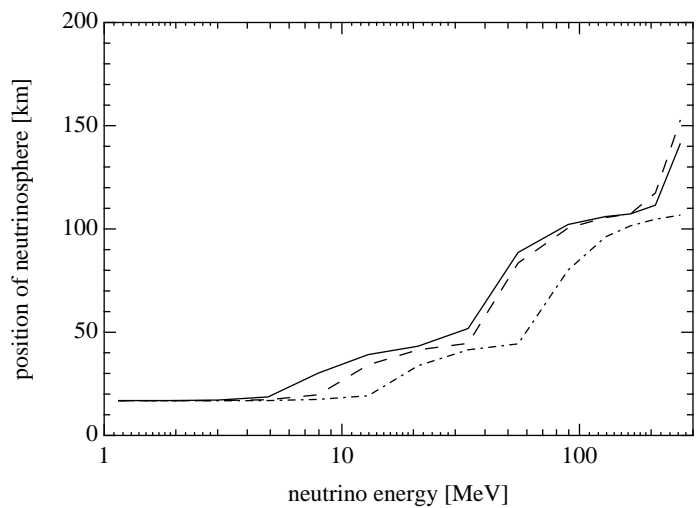}
\caption{The positions of neutrinosphere 
for $\nu_e$, $\bar{\nu}_e$ and $\nu_{\mu/\tau}$ 
are shown as a function of neutrino energy 
by solid, dashed and dot-dashed lines, respectively, 
at t$_{pb}$=1 sec for model SH.}
\label{fig:sh-tpb1000-rnusph}
\end{figure}

The diffusion of neutrinos trapped inside the proto-neutron star 
is actually slow and the emission of neutrinos is mainly driven  
by the accretion.  
In Fig. \ref{fig:sh-tpb1000-lumi}, we show the profiles of neutrino 
luminosities at t$_{pb}$=1 sec in model SH as a function of radius, 
for example.  
The luminosities increase stepwise at $\sim$20km, 
which corresponds to the outer edge (surface) of proto-neutron star.  
This is because 
the cooling of material is maximum there producing neutrinos 
by the electron and positron captures on nucleons and 
by the pair (electron and positons) annihilation processes.  
The accreting matter falls down to the surface 
after being heated by shock wave and compression, and then 
cools down by emitting neutrinos while it settles down 
hydrostatically.  
In order to demonstrate the neutrino emitting region, 
we show, in Fig. \ref{fig:sh-tpb1000-rnusph}, the positions of 
neutrinosphere at t$_{pb}$=1 sec in model SH 
as a function of neutrino energy. 
We define the neutrinosphere where the optical depth becomes 2/3 
for the neutrino with a specific energy.  
The neutrinospheres are located at 20$\sim$200 km, 
where the accreting matter cools down by emitting neutrinos.
Some of the high energy neutrinos are reabsorbed by falling 
material and are not free-streaming up to $\sim$200 km.
It is to be noted that 
the neutrinosphere for $\nu_{\mu/\tau}$'s is located 
in the innermost part while the ones for $\nu_e$ and $\bar{\nu}_e$ 
in the outer part.  
This is because $\nu_{\mu/\tau}$'s interact only through 
neutral currents whereas $\nu_e$'s and $\bar{\nu}_e$'s interact 
through both neutral and charged currents.  
This difference leads to the hierarchy of the average energies 
for different neutrino species that we will see in neutrino signals, 
since the temperature becomes higher as one goes inwards.  

\subsection{Neutrino signals}\label{sec:neutrino}
\begin{figure}
\epsscale{1.1}
\plottwo{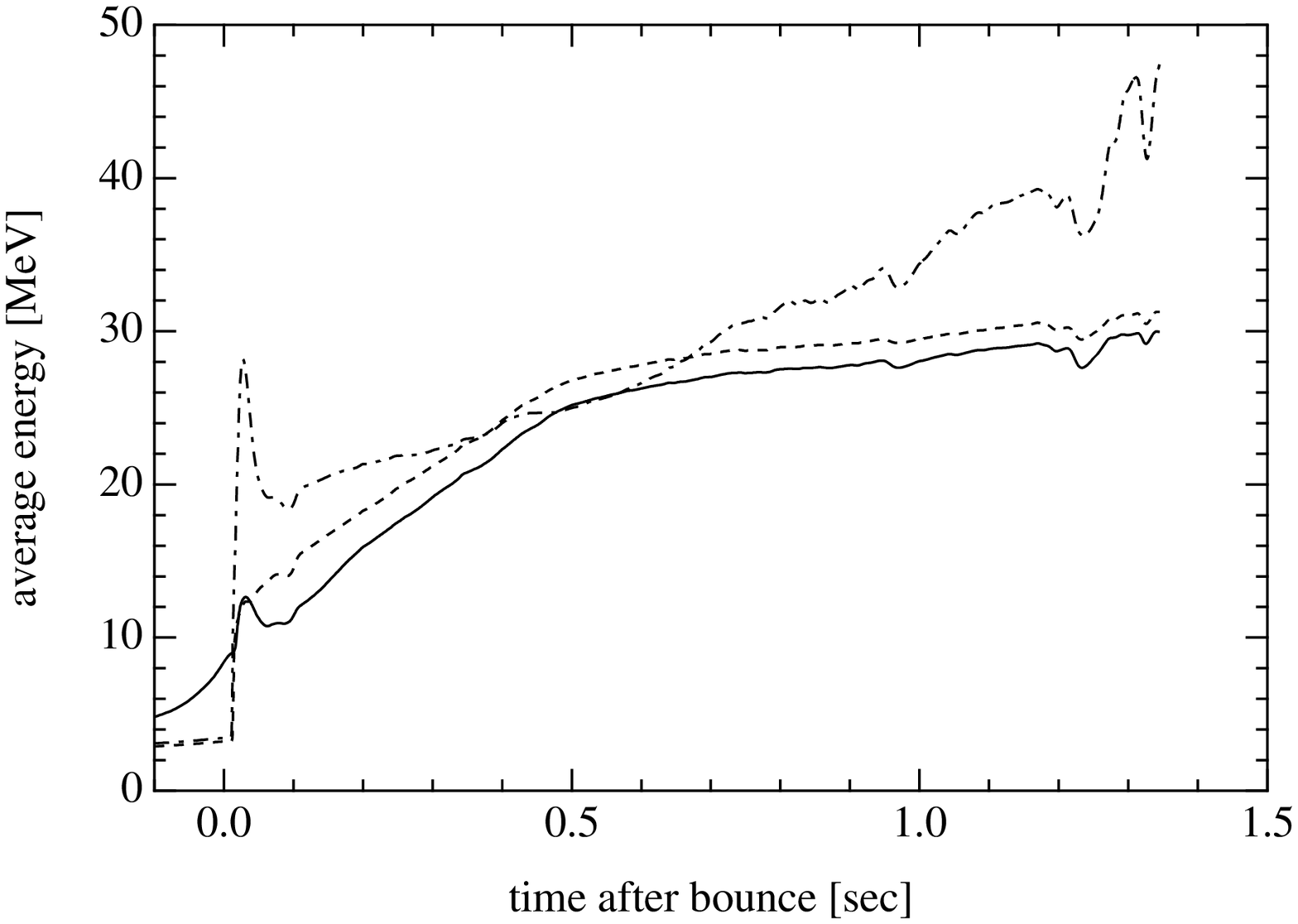}{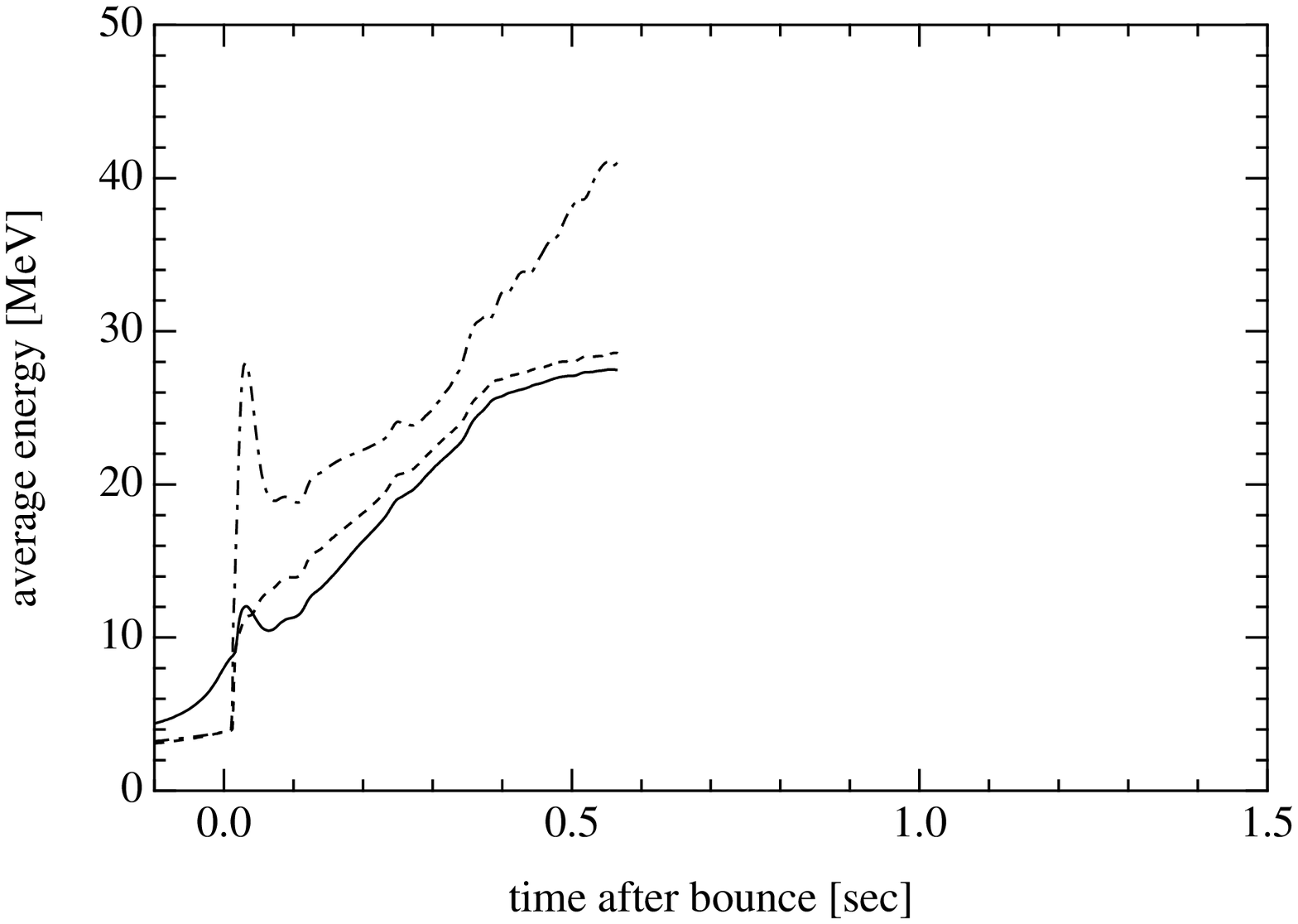}
\caption{Average energies 
of $\nu_e$ (solid), $\bar{\nu}_e$ (dashed) and $\nu_{\mu/\tau}$ (dot-dashed) 
as a function of time (t$_{pb}$) in models SH (left) and LS (right).}
\label{fig:eave}
\end{figure}

\begin{figure}
\epsscale{1.1}
\plottwo{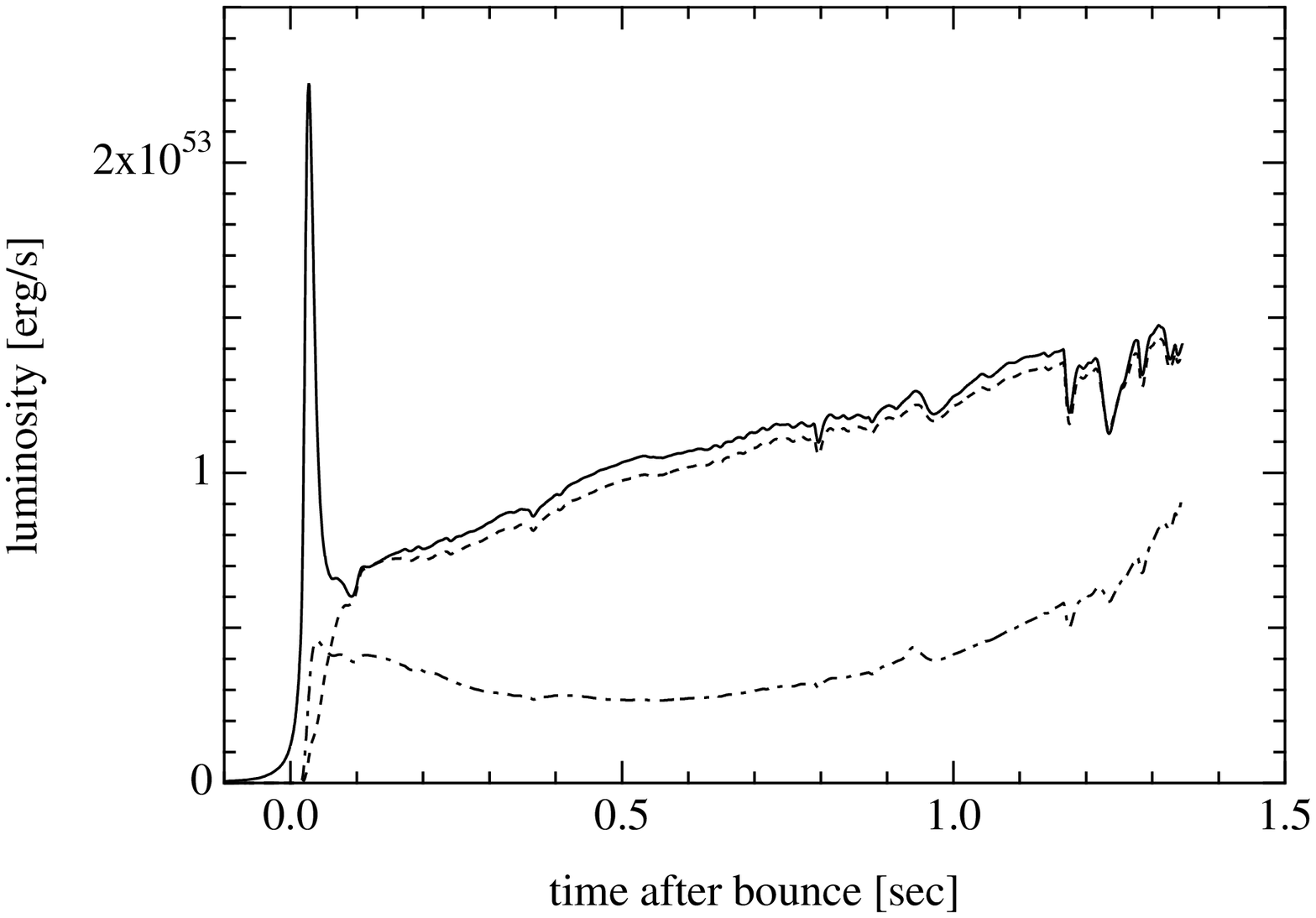}{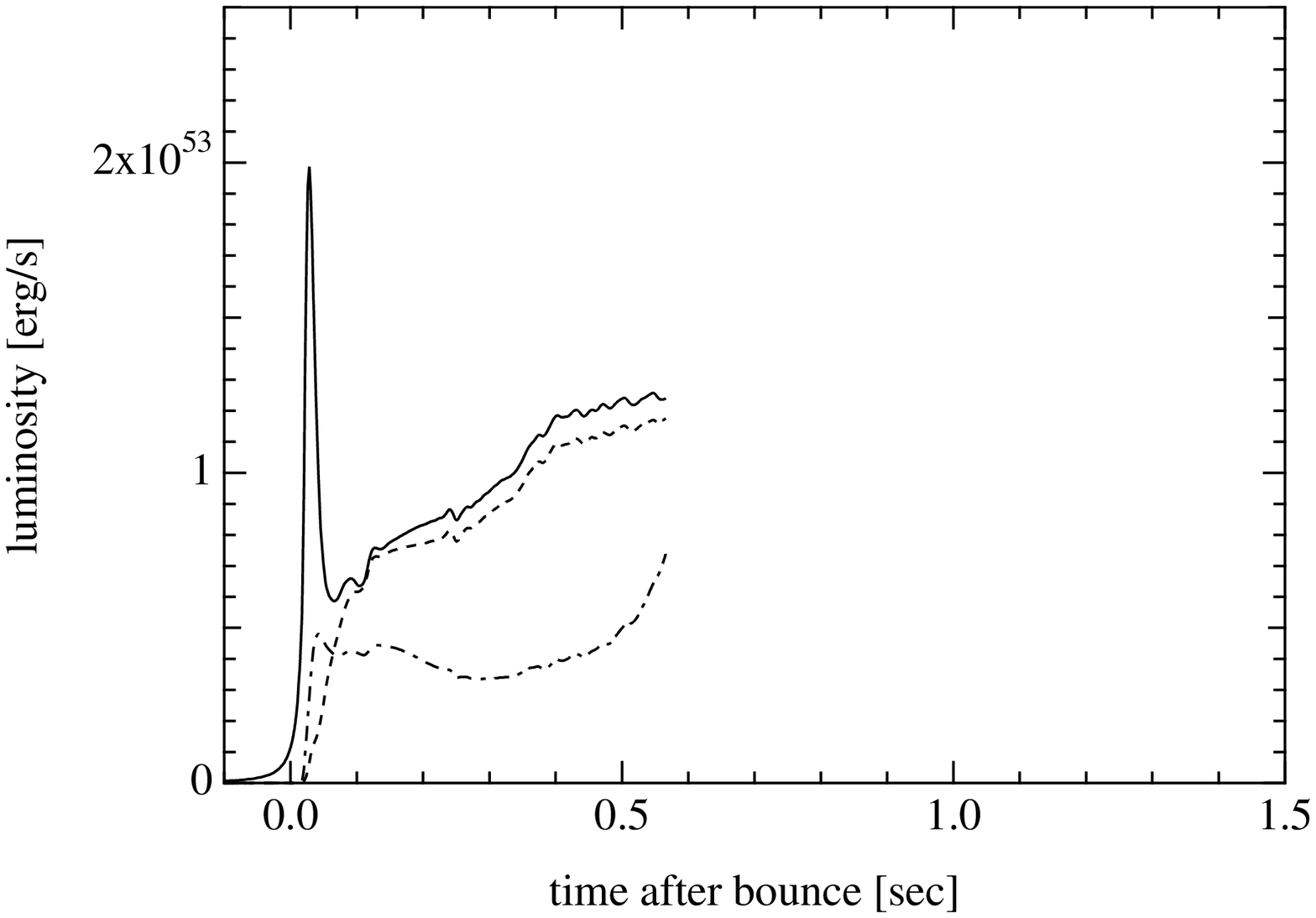}
\caption{Luminosities 
of $\nu_e$ (solid), $\bar{\nu}_e$ (dashed) and $\nu_{\mu/\tau}$ (dot-dashed) 
as a function of time (t$_{pb}$) in models SH (left) and LS (right).}
\label{fig:lumi}
\end{figure}

In Figs. \ref{fig:eave} and \ref{fig:lumi}, we show the average 
energies and luminosities of $\nu_e$, $\bar{\nu}_e$ and $\nu_{\mu/\tau}$ 
as a function of time (t$_{pb}$) for the two models.  
These quantities are the ones measured at the outermost grid point 
($\sim$6000 km).  
The average energy presented here is defined 
by the rms value, $< E_{\nu}^{2} >^{\frac{1}{2}}$.
The plots for $\bar\nu_{\mu/\tau}$ are not shown since they 
show no significant difference from the ones for $\nu_{\mu/\tau}$.  
We will only discuss $\nu_{\mu/\tau}$ and will not mention 
$\bar\nu_{\mu/\tau}$ hereafter.  

Around the core bounce ($t_{pb}=-0.1 \sim 0.1$ s), 
the time profiles of average energies and luminosities 
are similar to those for ordinary supernovae.  
There is a distinctive peak due to the neutronization burst 
in the luminosity of $\nu_{e}$.  
The rise of the luminosities of $\bar{\nu}_e$ and $\nu_{\mu/\tau}$ 
right after bounce occurs owing to the thermal production of neutrinos.  
The peak of average energies around the core bounce 
appears as a result of heating by the passage of shock wave.  
The behaviors in two models SH and LS are similar to each other 
up to $t_{pb} \sim 0.1$ s.  

After that, the average energies increase toward the black hole 
formation, reflecting the temperature increase.  
This tendency is more evident in model LS since the contraction 
of proto-neutron star is faster and the resulting temperature 
is higher.  
The average energy of $\nu_{\mu/\tau}$ increases most prominently 
among three species, having the neutrinosphere at the innermost.  
The average energies of $\nu_{e}$ and $\bar{\nu}_e$ are 
rather close to each other because 
their neutrinospheres locate at similar positions,
which are determined by charged current 
reactions on mixture of neutrons and protons.
It is remarkable to see that the average energies rise 
continuously up to the end and the increase amounts to 
a factor of two to three from the value right after the bounce.  
This behavior of neutrino emission is different from the one 
seen usually in the numerical simulations of ordinary supernovae.  
In the latter, the proto-neutron star is 
more static and does not contract so much 
because the accretion ceases soon after the bounce.  
Therefore, the increase of average energies after bounce 
is a clear signal that tells us the evolution toward the 
black hole formation.  

The luminosities after the neutronization burst also increase 
toward the black hole formation.  
This is in accord with the increase of accretion luminosity, 
especially for $\nu_{e}$ and $\bar{\nu}_e$.  
The accretion luminosity is proportional to $GM\dot{M}/r$, 
where $M$ and $r$ denote the mass and radius, respectively, of 
the proto-neutron star discussed in $\S$ \ref{sec:pns}.  
While the accretion rate, $\dot{M}$, stays roughly constant, 
the radius decreases dramatically 
like the shock positions in Figs. \ref{fig:traj-sh} and \ref{fig:traj-ls} 
toward the end.  
The actual neutrino luminosities, which are a portion of 
the available energy obtained by the accretion, 
depend on the cooling processes.  
The cooling by $\nu_{e}$ and $\bar{\nu}_e$ emissions 
proceeds through the electron and positron captures 
on nucleons in hot accreting matter.  
Since the matter is non-degenerate and the electron 
fraction is $\sim$0.5 containing 
both neutrons and protons equally, 
the $\nu_{e}$ and $\bar{\nu}_e$ luminosities are 
nearly equal.  
The $\nu_{\mu/\tau}$ luminosities are determined by 
the pair process (annihilation of electron-positron pairs), and, 
therefore, reflect the temperature change at the neutrinosphere 
for $\nu_{\mu/\tau}$'s.  

It is to be noted that the computations of neutrino burst  
are terminated 
at the formation of apparent horizon in this study.  
After this moment, the neutrinosphere will be swallowed 
by the horizon in a fraction of millisecond and the major neutrino 
emissions cease at this point.  
Since neutrinos emitted just at the neutrinosphere 
travel over $\sim$6000 km to the observing point 
set at the outer boundary of the computation domain, 
the neutrino signals last for $\sim$20 msec more, 
in reality, from the end point in the current figures.  
The last part of neutrino signals will be affected more clearly 
by general relativity, e.g. as the energy red shift \citep{bau96b}.  
Such detailed information of neutrino spectra is 
valuable to examine general relativistic effects 
and to constrain neutrino masses \citep{bea01}.  
In order to predict the final moment of neutrino emission, however, 
one needs to implement a scheme to avoid both coordinate- and 
real singularities \citep{bau95}, 
which will be postponed to the future work.  
The scope of the current study is 
to reveal the general feature of the evolution 
from the massive progenitor to the black hole.  
It is the difference in the duration of proto-neutron star era 
and associated neutrino bursts 
that we aim to clarify using the two realistic equations of 
state of dense matter.

\begin{figure}
\epsscale{1.0}
\plottwo{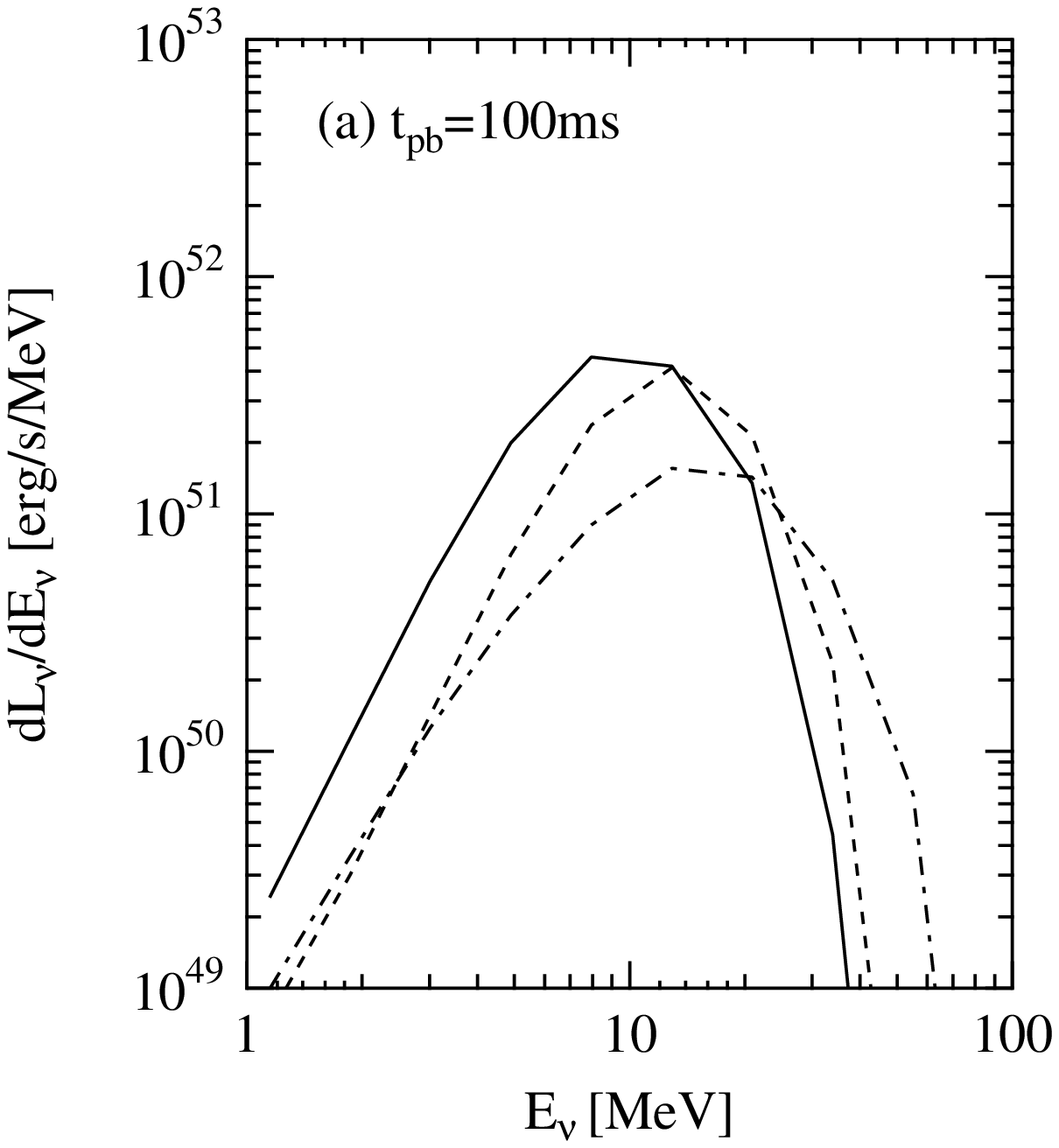}{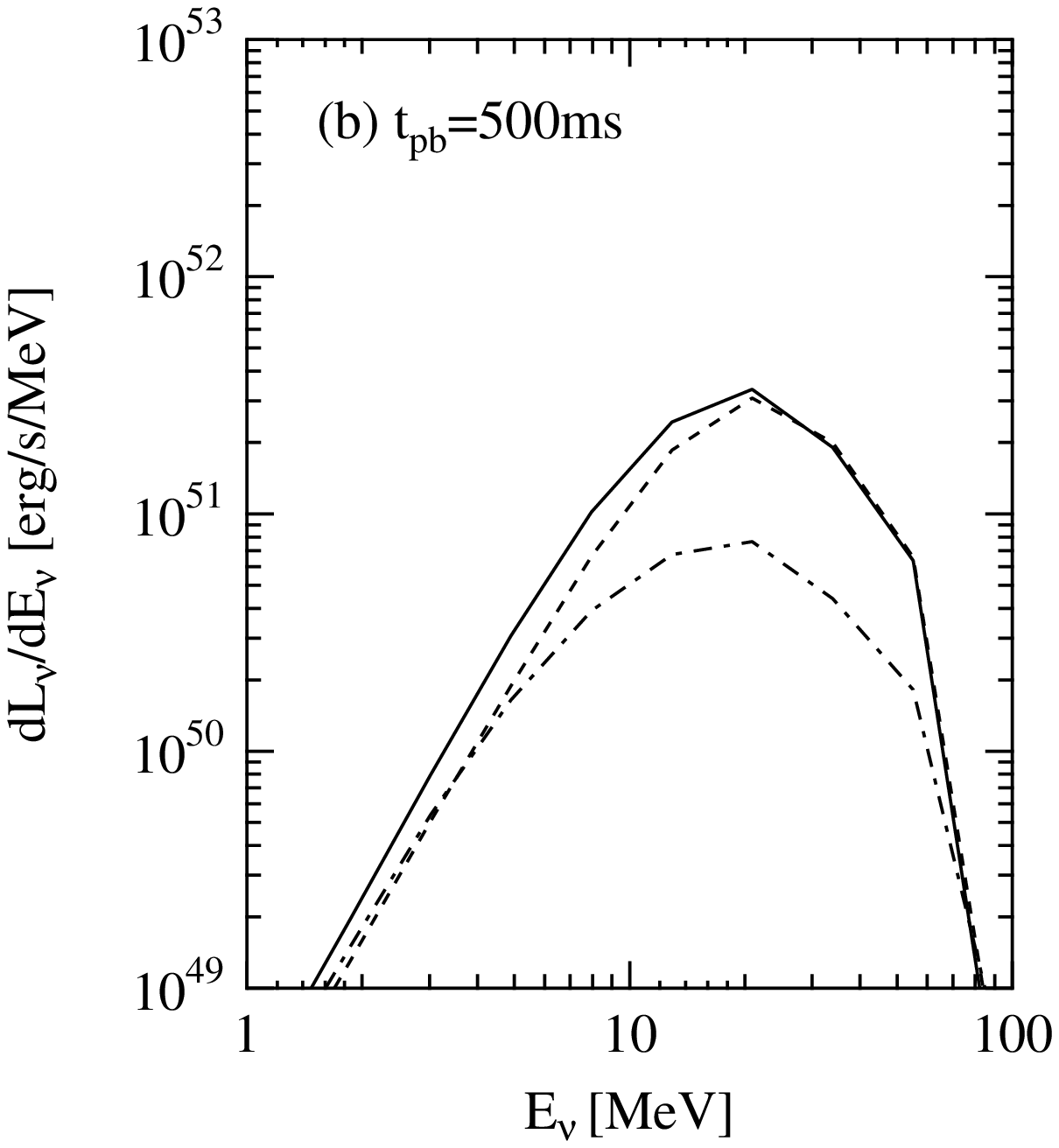}\\
\plottwo{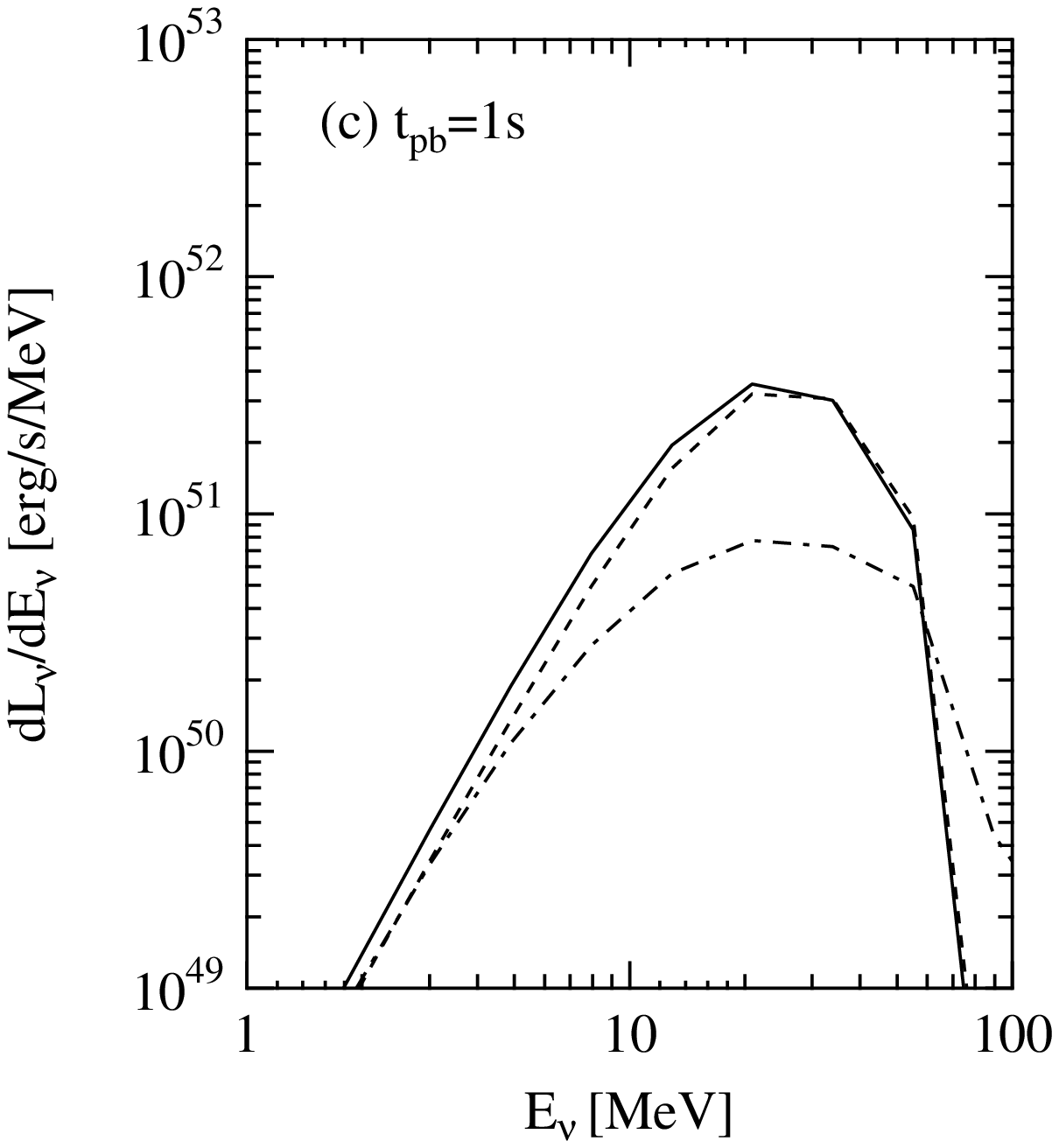}{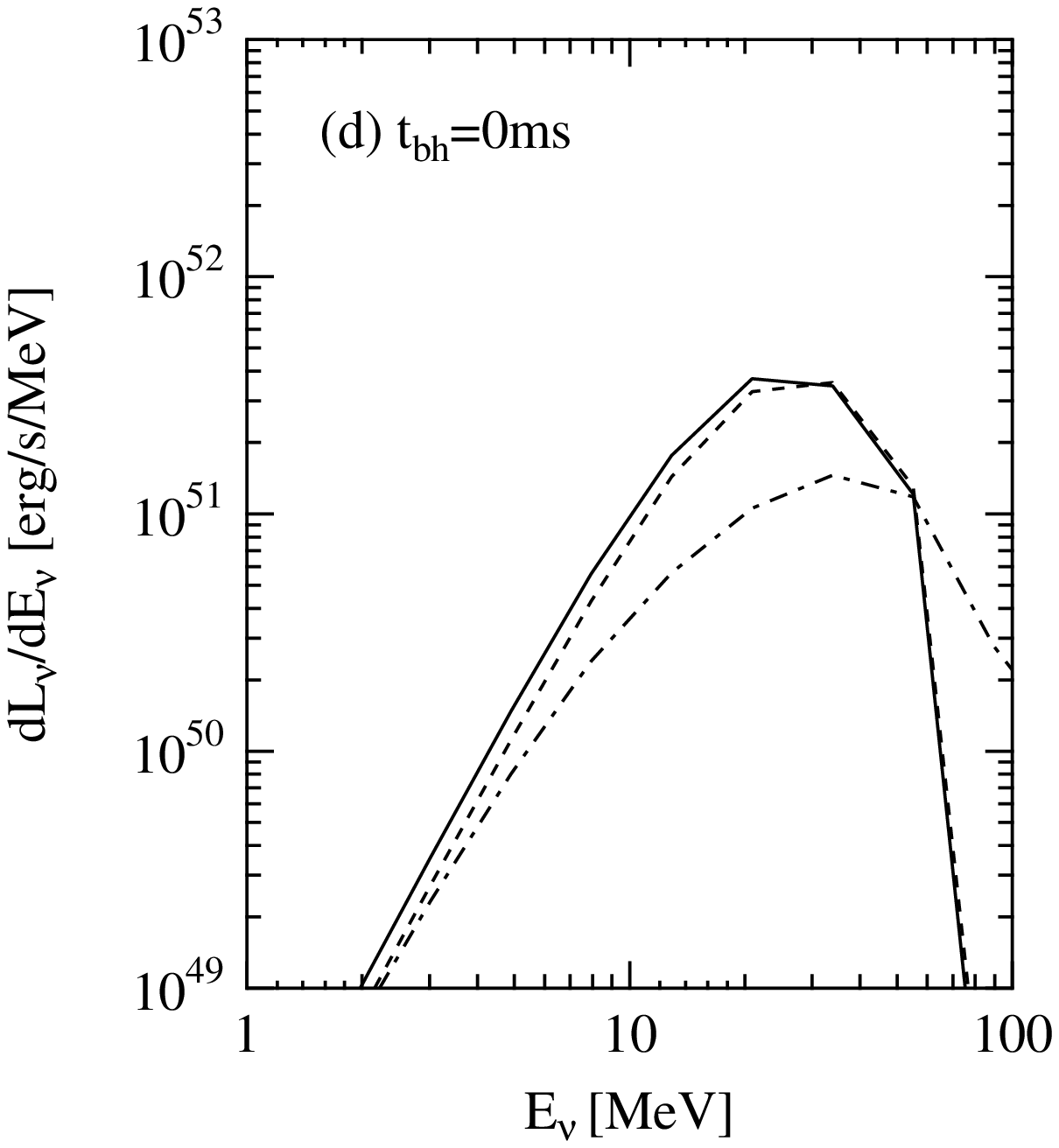}
\caption{Energy spectra are shown 
as a function of neutrino energy for 
$\nu_e$, $\bar{\nu}_e$ and $\nu_{\mu/\tau}$ 
by solid, dashed and dot-dashed lines, respectively, 
at t$_{pb}$=100 msec, 500 msec, 1 sec and t$_{bh}$=0 for model SH.}
\label{fig:sh-spect}
\end{figure}

In Fig. \ref{fig:sh-spect}, 
we show the time evolutions of energy spectra 
at the outermost grid point for model SH.  
It is apparent that the spectra become harder during the 
evolution toward the black hole formation as the average 
energies and the luminosities evolve 
(see also Figs. \ref{fig:eave} and \ref{fig:lumi}).  
At $t_{pb}$=100 ms, the peak of spectrum is located around 
10$\sim$20 MeV having the same hierarchy as in the average energies. 
The peak is shifted to higher energies ($\sim$20 MeV) and 
the spectral shapes become similar among three neutrino species 
at $t_{pb}$=500 ms.  
The spectra for $\nu_{e}$ and $\bar{\nu}_e$ are similar to 
each other due to the comparable production mechanisms as discussed above.  
The spectra for $\nu_{\mu/\tau}$'s become rapidly harder 
further toward $t_{bh}$=0 reflecting higher temperatures of 
the central object.  

\begin{figure}
\epsscale{1.0}
\plotone{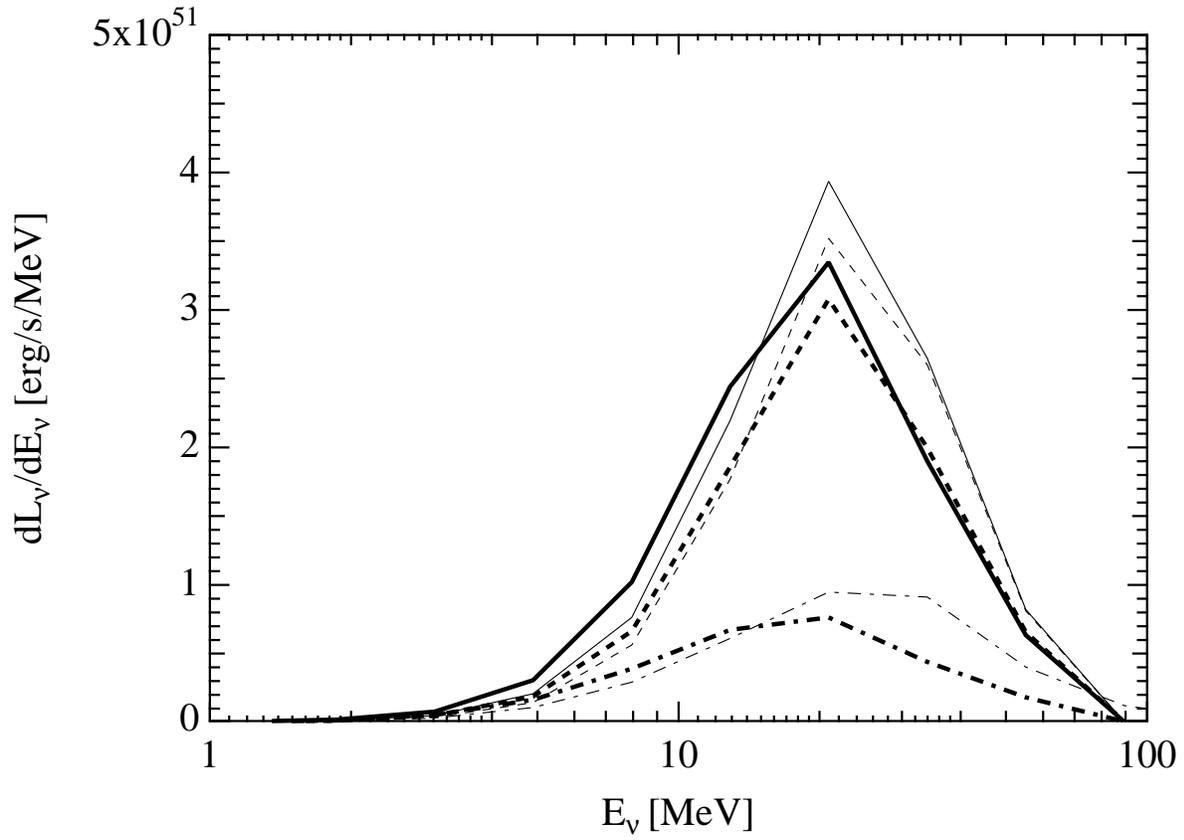}
\caption{Energy spectra are shown 
as a function of neutrino energy for 
$\nu_e$, $\bar{\nu}_e$ and $\nu_{\mu/\tau}$ 
by solid, dashed and dot-dashed lines, respectively, 
at t$_{pb}$=500 msec for model SH (thick) and LS (thin).}
\label{fig:spect-tpb500}
\end{figure}

We compare the energy spectra for models SH and LS 
and found that the difference is minor 
as long as we compare them at the same timing after bounce.  
The spectra at $t_{pb}$=100 ms for LS is quite similar 
to the ones shown in (a) of Fig. \ref{fig:sh-spect}.  
We show the comparison at $t_{pb}$=500 ms in Fig. \ref{fig:spect-tpb500}.  
The luminosities for model LS are slightly higher 
than those for model SH.  
The $\nu_{\mu/\tau}$ spectra for model LS are 
harder than those for model SH, while 
the $\nu_{e}$ and $\bar{\nu}_e$ spectra look similar 
for both models.  
These differences arise from the fact that 
the proto-neutron star in model LS 
at $t_{pb}$=500 ms is already close to the state 
of the maximum mass whereas the model SH waits 
for the further evolution.  


\section{Implications for EOS studies}\label{sec:EOS-implications}
\begin{figure}
\epsscale{1.1}
\plottwo{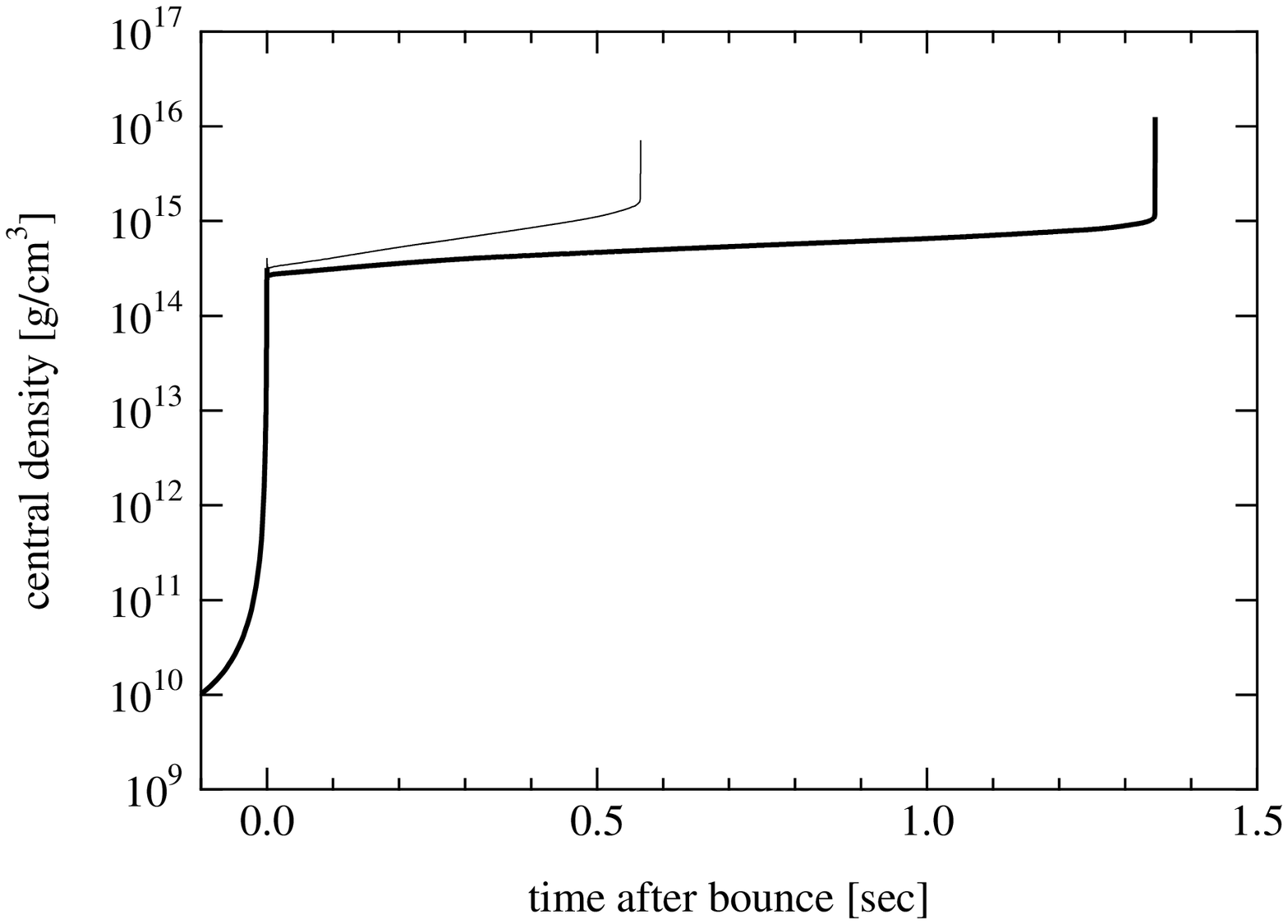}{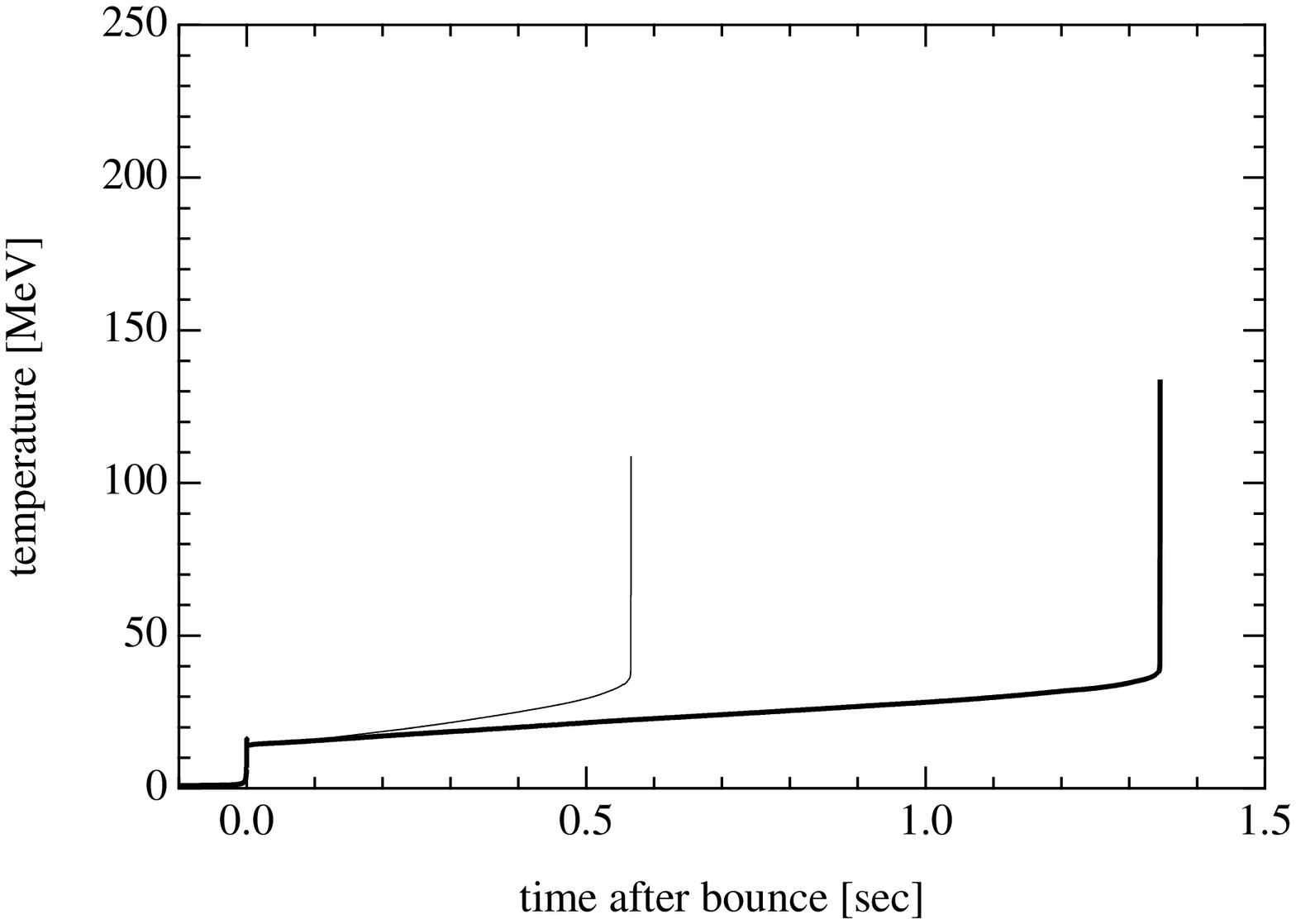}
\caption{Density (left) and temperature (right) 
along the trajectory at center are shown 
as a function of time (t$_{pb}$) for models SH (thick) and LS (thin).}
\label{fig:mb00}
\end{figure}

\begin{figure}
\epsscale{1.1}
\plottwo{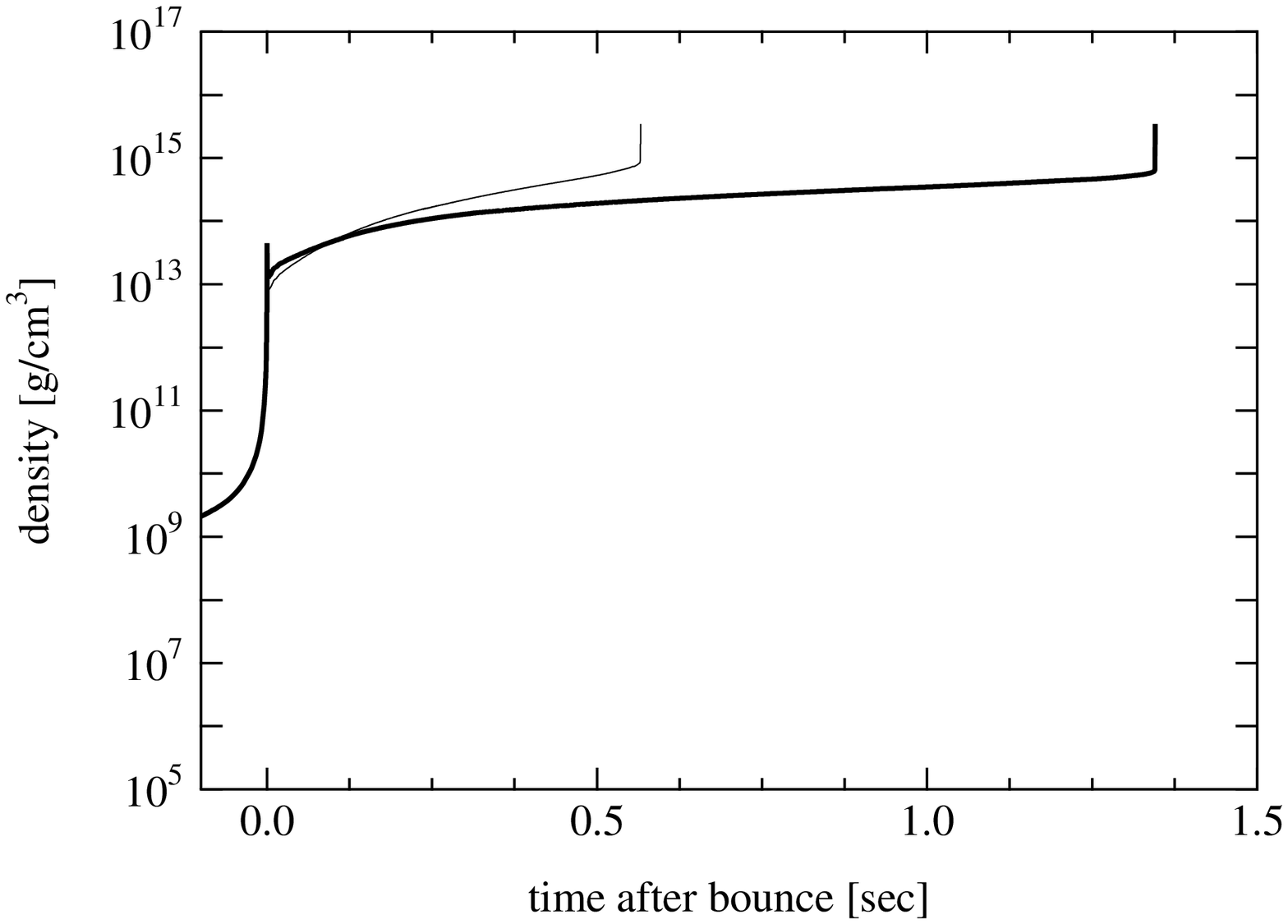}{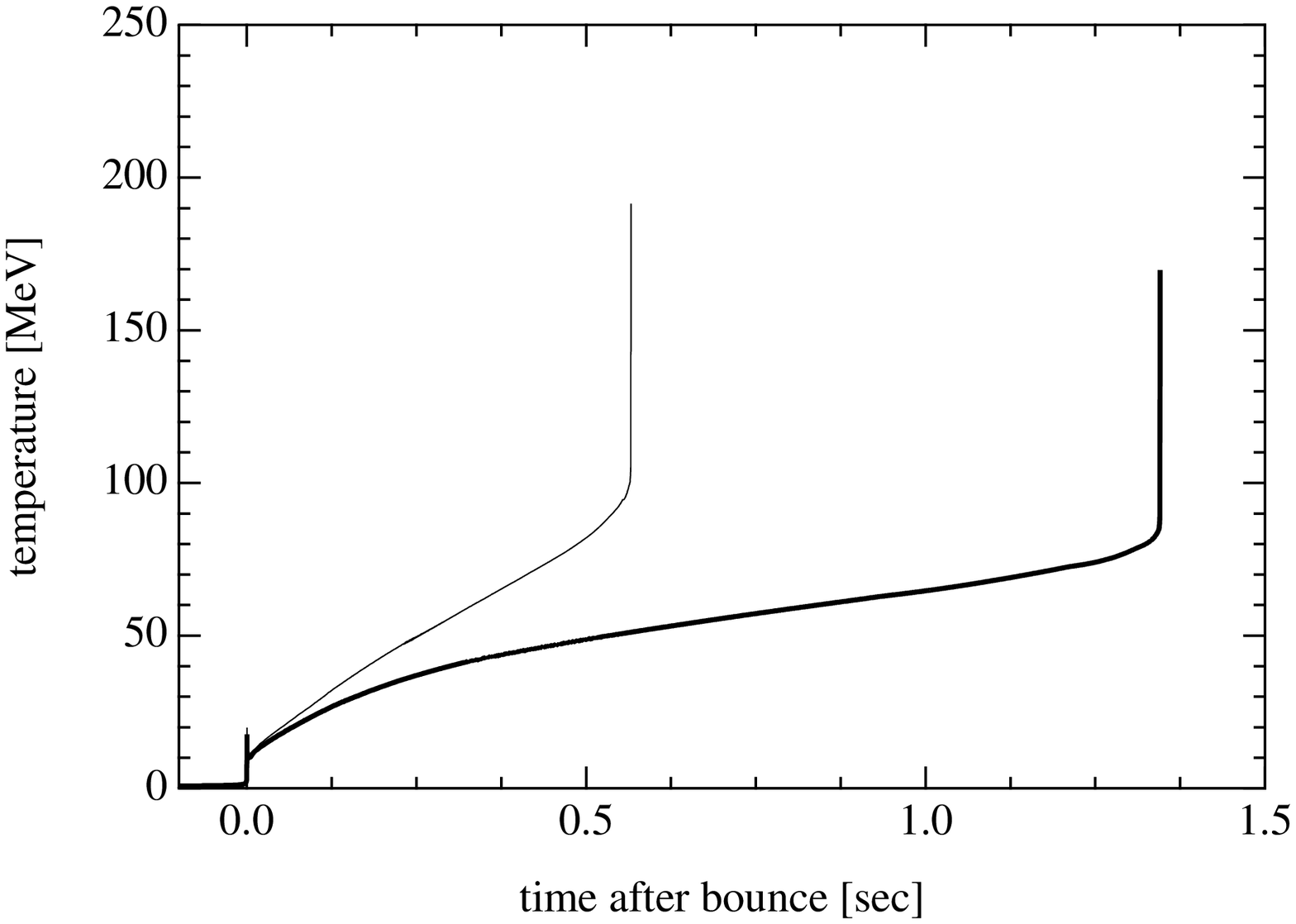}
\caption{Density (left) and temperature (right) 
along the trajectory at M$_{b}$=0.6M$_{\odot}$ are shown 
as a function of time (t$_{pb}$) for models SH (thick) and LS (thin).}
\label{fig:mb06}
\end{figure}

Here we describe the temperature and density regimes 
that the stellar matter experiences during the evolution 
to the black hole formation and 
discuss their implications for the further development of EOS.  
In Figs. \ref{fig:mb00} and \ref{fig:mb06}, we plot the density 
and temperature of matter along the trajectory for a fixed baryon 
mass coordinate as a function of time after bounce for the two models.
At the center (Fig. \ref{fig:mb00}), the density increases quickly 
before the core bounce and gradually thereafter toward the re-collapse.  
By the time of re-collapse, the density reaches 10$^{15}$ g/cm$^{3}$, 
which is 3 times the normal nuclear matter density.  
The temperature exceeds 10 MeV at bounce, and gradually 
increases beyond 30 MeV toward the re-collapse.  
The final sharp increases we can see in figures are 
due to the dynamical collapse to the black hole 
within a millisecond.  
We remark that the behavior before the sharp increase 
is important to determine 
the trigger of the gravitational collapse at the maximum mass 
during the evolution of proto-neutron stars.  
In Fig. \ref{fig:mb06}, we show the case for 0.6M$_{\odot}$ 
in baryon mass coordinate, where the entropy peak 
is formed by the passage of shock wave.  
In this case, the temperature increases fast and 
becomes much higher than at the center.  
The temperature reaches 80 MeV for SH (90 MeV for LS) 
before the re-collapse and the density exceeds twice 
the normal nuclear matter density.  
These high density and temperature regimes are important to 
study the rapid collapse of proto-neutron stars.  

As we described in $\S$ \ref{sec:EOS}, 
we adopt the equations of state with only the nucleonic degree of freedom 
in the current study.  
For high densities and temperatures we have seen above, 
more massive baryons including strangeness may appear 
and the quark degree of freedom becomes essential 
to describe the dense and hot matter.
An emergence of new degrees of freedom will lead to the softening 
of equation of state and, therefore, it makes the maximum mass of 
proto-neutron star smaller and the re-collapse earlier.  
In this sense, the current result gives the maximum duration of 
proto-neutron star era.  
Still, it is astonishing that we see a large difference 
even within the nucleonic EOSs.  
The current results are also the basis to explore 
when the exotic phases of dense matter appear during the evolution.  
The determination of the end point of neutrino burst will be 
important to put a 
constraint on the phase transitions of dense matter.  
A shorter burst than the current prediction may suggest 
the existence of new phases on the density-temperature trajectories 
studied here.  
It is certainly important to constrain the nucleonic EOS 
further as a firm basis as well as to perform numerical simulations 
with the extended EOS table with exotic phases.  
The systematic study on the hyperon and/or quark EOSs covering 
a wide range of environment and on the associated neutrino reactions is 
currently under way.  

In the current study, we have used the incompressibility of 180 MeV 
among the three choices in LS-EOS.  
If we use LS-EOS with the higher incompressibility of 220 MeV, 
the duration of neutrino signal will become longer 
than in the current case of 180 MeV and 
the difference between LS-EOS and SH-EOS will be smaller 
than for the results we have shown.  
Since the characteristics of EOS depends not only on the incompressibility 
but also on the symmetry energy, adiabatic index and other factors 
at high densities, it would be interesting to extract the dependence 
on the incompressibility alone by numerical simulations 
using LS-EOS with the higher incompressibility.  


\section{Summary}\label{sec:summary}
We study the gravitational collapse of a non-rotating massive star 
of 40 M$_\odot$ 
resulting in the black hole formation.  
Adopting the two different sets of realistic equation of state (EOS) of dense matter, 
we perform the numerical simulations of general relativistic $\nu$-radiation hydrodynamics.  
We reveal the influence of EOS on the core bounce, the shock propagation, 
the formation of proto-neutron star and the ensuing re-collapse to black hole 
together with neutrino emissions during the evolution.
Following the core bounce and shock stall, the accretion of material 
from outer layers causes the rapid increase 
of the mass of nascent proto-neutron star.
Accordingly, the proto-neutron star immediately starts 
shrinking within 100 msec after bounce.
The speed of contraction depends predominantly on the adopted EOS, 
and the density and temperature of the proto-neutron star become very high 
as compared with those for 
ordinary proto-neutron stars in successful supernovae.  
We find that the duration of proto-neutron star era is clearly different 
depending on the EOS, which determines the maximum mass of proto-neutron stars.  
Due to the gravitational instability beyond the maximum mass, 
the fat proto-neutron star collapses again at 0.56 s and 1.34 s 
after bounce for the models with LS-EOS and SH-EOS, respectively.  
The apparent horizon is formed within $\sim$1 msec.
The black hole is formed much more quickly 
after the initial collapse of massive star 
than in the delayed scenarios.  
The resulting $\nu$ emissions during the dynamical evolution 
are distinctive as compared with the neutrino signals 
not only from ordinary supernovae
but also from the delayed collapse models.  
The ever increasing average energies of neutrinos and the sudden termination 
of neutrino signals are the clear evidence of the rapid black hole formation 
from massive stars considered here.  
Although the increasing feature of energy can be seen also 
in some models of successful explosions, it is more robust in this case 
due to the lasting intense accretion.  
In addition, the duration of $\nu$ emissions can be used to constrain 
the stiffness of EOS at high density and temperature.  
A softer EOS or an early occurrence of the phase transition to exotic phases 
may lead to a shorter neutrino burst.  
Further studies concerning the dependences on progenitor models and 
the EOS with hyperon or quarks are necessary to provide the templates 
for the neutrino signal of black hole formation
and will be reported in separate papers.
It would be interesting to study the black hole formation 
from rotating massive stars since the rotation will modify 
the formation time and the associated neutrino signal, 
which may contain the contributions from accretion disks.  
A terrestrial detection of neutrino bursts from this type of collapse 
of massive stars in future will reveal the rapid black hole formation 
and will put a new constraint on the EOS at extreme conditions.  

\acknowledgments
The authors are grateful to H. Shen, K. Oyamatsu, 
A. Ohnishi, C. Ishizuka and H. Toki 
for the collaborations on the tables of equation of state 
for supernova simulations.  
K. S. is grateful to T. Maruyama and S. Chiba 
for the arrangement of computing resources at JAEA.  
We thank M. Liebendorfer, K. Nakazato, S. Fujimoto 
and T. Kajino for fruitful discussions and keen comments.
The numerical simulations were performed 
at NAO (wks06a, iks13a), JAEA and YITP.
This work is partially supported by the Grants-in-Aid for the 
Scientific Research (15540243, 15740160, 18540291, 18540295) 
of the MEXT of Japan, Academic Frontier Project and KEK LSSP (06-02).  

\clearpage

\clearpage

\clearpage



\clearpage

\clearpage

\clearpage

\clearpage

\clearpage

\clearpage

\clearpage

\clearpage

\clearpage

\end{document}